\documentclass{pasa}%

\usepackage{graphicx}
\usepackage{amsmath}
\usepackage{amssymb}
\usepackage{subfigure}
\usepackage{sidecap}

\usepackage{aas_macros}
\usepackage{xcolor}
\usepackage{hyperref}
\hypersetup{colorlinks,citecolor=blue,linkcolor=blue,urlcolor=blue}

\providecommand{\aap}{Astron.Astrophys.}
\providecommand{\apjs}{Astrophys.J.Suppl.}
\providecommand{\apjl}{Astrophys.J.}
\providecommand{\physrep}{Phys. Rep.}
\providecommand{\mnras}{Mon. Not. Roy. Astron. Soc.}
\providecommand{\npa}{Nucl. Phys. A}

\providecommand{\epj}{Eur. Phys. J.}
\providecommand{\epja}{Eur. Phys. J. A}
\providecommand{\rmp}{Rev.~Mod.~Phys.}
\providecommand{\prc}{Phys. Rev. C}
\providecommand{\prd}{Phys. Rev. D}
\providecommand{\pr}{Phys. Rev.}

\title{The state of matter in simulations of core-collapse supernovae -- Reflections and recent developments}

\author[T. Fischer et al.]{
Tobias Fischer$^1$\thanks{email: fischer@ift.uni.wroc.pl},
Niels-Uwe Bastian$^{1}$,
David Blaschke$^{1,2,3}$,
Mateusz Cierniak$^{1}$,
Matthias Hempel$^4$,
Thomas Kl{\"a}hn$^5$,
Gabriel Mart{\'i}nez-Pinedo$^{6,7}$,
William G. Newton$^8$,
Gerd R{\"o}pke$^{3,9}$
and
Stefan Typel$^{7,6}$
\\
\affil{$^1$ Institute of Theoretical Physics, University of Wroclaw, Pl. M. Borna 9, 50-204 Wroclaw, Poland}%
\affil{$^2$ Bogoliubov Laboratory for Theoretical Physics, Joint Institute for Nuclear Research, 141980 Dubna, Russia}
\affil{$^3$ National Research Nuclear University (MEPhI), Kashirskoe shosse 31, 115409 Moscow, Russia}
\affil{$^4$ Department of Physics, University of Basel, Klingelbergstrasse 82, 4058 Basel, Switzerland}
\affil{$^5$ Department of Physics and Astronomy, California State University Long Beach, 250 Bellflower Boulevard, Long Beach, California 90840-9505, USA}
\affil{$^6$ GSI Helmholtzzentrum f{\"u}r Schwerionenforschung, Planckstra{\ss}e 1, 64291 Darmstadt, Germany}
\affil{$^7$ Institut f{\"u}r Kernphysik, Technische Universit{\"a}t Darmstadt, Schlossgartenstra{\ss}e 9, 64289 Darmstadt, Germany}
\affil{$^8$ Department of Physics and Astronomy, Texas A\&M University-Commerce, Commerce, TX 75429, USA}
\affil{$^9$ Institut f{\"u}r Physik, Universit{\"a}t Rostock, Albert-Einstein Stra{\ss}e 23--24, 18059 Rostock, Germany}
}%

\jid{
95.85.Ry   
95.30.Qd,  
21.65.Qr   
}
\doi{10.1017/pas.\the\year.xxx}
\jyear{\the\year}

\begin{document}

\begin{frontmatter}
\maketitle

\begin{abstract}
In this review article we discuss selected developments regarding the role of the equation of state (EOS) in simulations of core-collapse supernovae. There are no first-principle calculations of the state of matter under supernova conditions since a wide range of conditions is covered, in terms of density, temperature and isospin asymmetry. Instead, model EOS are commonly employed in supernova studies. These can be divided into regimes with intrinsically different degrees of freedom: heavy nuclei at low temperatures, inhomogeneous nuclear matter where light and heavy nuclei coexist together with unbound nucleons, and the transition to homogeneous matter at high densities and temperatures. In this article we discuss each of these phases with particular view on their role in supernova simulations.
\end{abstract}

\end{frontmatter}

\section{INTRODUCTION }
\label{sec:intro}

\begin{figure*}[btp!]
\hspace{-5mm}
\subfigure[Temperature-density domain of a supernova evolution.]{
\includegraphics[width=1.15\columnwidth]{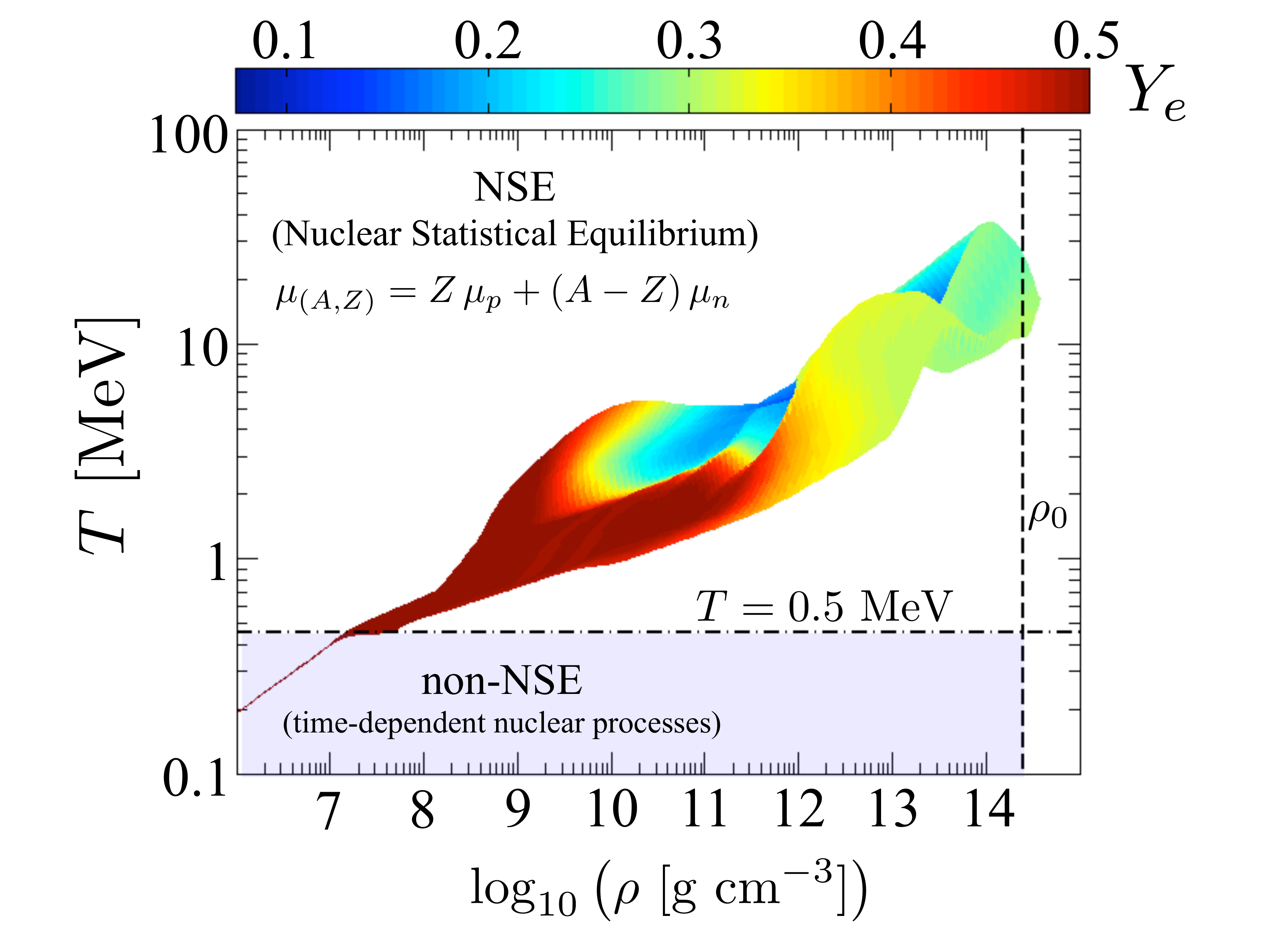}\label{fig:eos}}
\hspace{-5mm}
\subfigure[Space-time diagram of the supernova evolution.]{
\includegraphics[width=1.10\columnwidth]{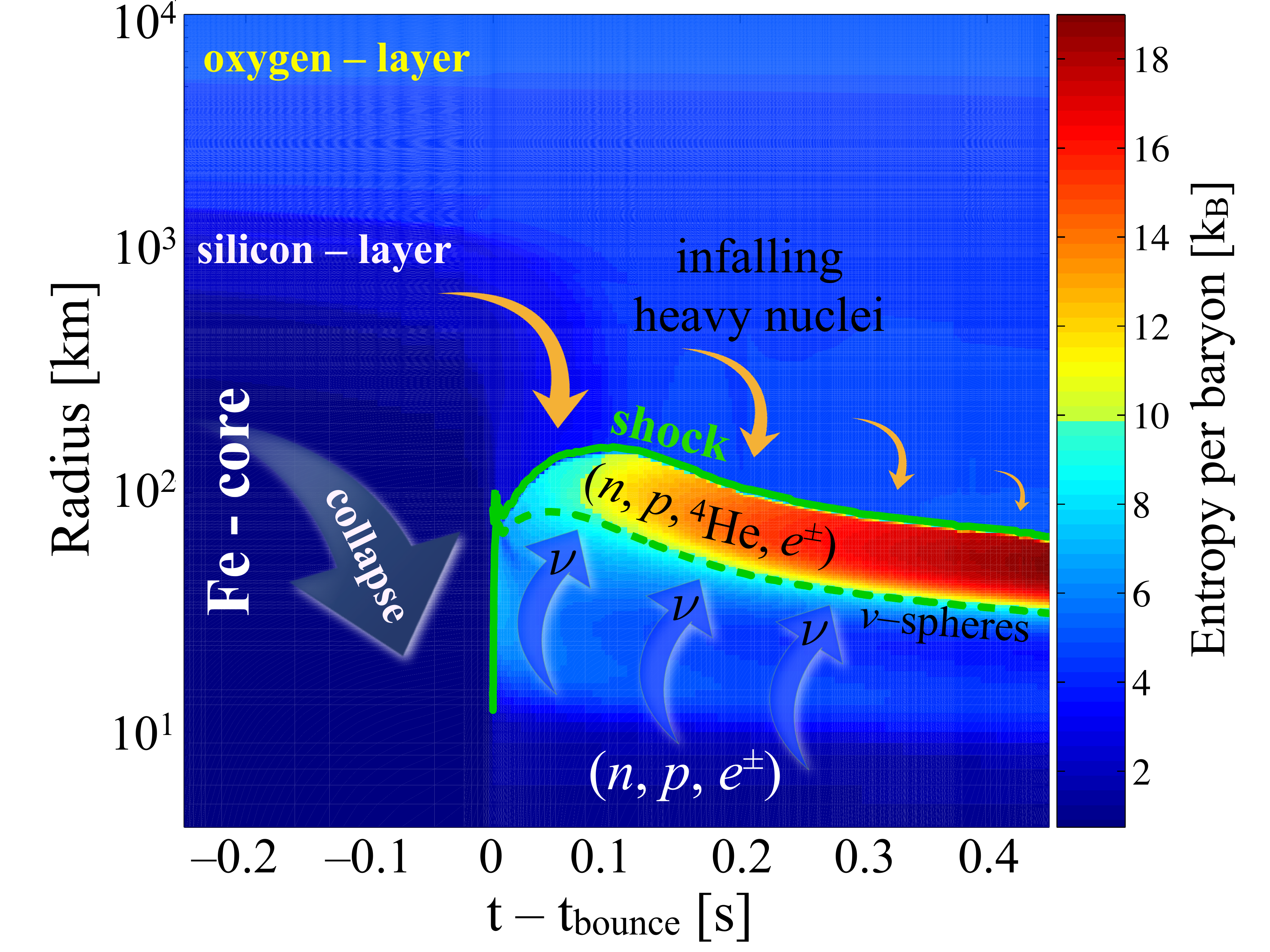}\label{fig:shellplot}}
\caption{Supernova phase diagram (color-coding is due to the electron fraction $Y_e$) in graph~(a) and space-time diagram of the supernova evolution (color-coding is due to the entropy per baryon) in graph~(b), both obtained from the spherically-symmetric core-collapse supernova simulation of the massive progenitor star of 18~M$_\odot$ published in \cite{Fischer:2016a}.}
\label{fig1}
\end{figure*}

The state of matter at the interior of core-collapse supernovae as well as in the core of proto-neutron stars (PNS) can reach extreme conditions, in terms of temperatures up to several $10^{11}$~K ($1.16\times10^{10}~\rm K\simeq 1$~MeV), densities in excess of normal nuclear matter density and large isospin asymmetry. The associated supernova phase diagram, i.e. the thermodynamic conditions obtained during a core-collapse supernova, is shown in Fig.~\ref{fig:eos} for a selected example simulation. Central densities and maximum temperatures may vary on the order of 10--20\% depending on the stellar progenitor. The EOS for supernova simulations must cover such an extended three-dimensional domain illustrated in Fig.~\ref{fig:eos}, where presently first-principle EOS are not available. Instead, model EOS are being developed for supernova studies. These combine several domains with different degrees of freedom, e.g., heavy nuclei at low temperatures, inhomogeneous nuclear matter composed of light and heavy nuclei together with unbound nucleons, and homogeneous matter at high temperatures and densities. In this article we reflect on the role of the EOS in core-collapse supernovae explored in spherically symmetric simulations. To this end accurate three-flavor Boltzmann neutrino transport, developed by \citet{Mezzacappa:1993gm,Mezzacappa:1993gn,Mezzacappa:1993gx}, is employed in the fully self-consistent general relativistic radiation-hydrodynamics framework of \citet{Liebendoerfer:2004}. In general, accurate neutrino transport is essential for the prediction of the neutrino signal for core-collapse supernova events, as was observed from SN1987A \citep[cf.][]{Bionta:1987qt,Hirata:1988ad} which marked a benchmark for current supernova modeling. The latter already confirmed that neutrinos from the next galactic event will be the potentially observable signal, from which we will learn not only details about the stellar explosion but also about the state of matter at the supernova interior which is hidden otherwise, e.g., for electromagnetic radiation by the stellar mantle. 

\subsection{Supernova phenomenology}
PNS are the central object of core-collapse supernovae. The latter are being triggered from the initial implosion of the stellar core of stars more massive than about 8~M$_\odot$. The core collapse proceeds until normal nuclear matter density is reached, when nuclei dissolve into homogeneous matter. The highly repulsive nature of the short-range nuclear force balances gravity such that the core {\em bounces} back with the formation of a hydrodynamics shock wave which propagates quickly out of the stellar core, as illustrated in Fig.~\ref{fig:shellplot} (thick solid green line). The shock stalls at around 100--200~km due to the continuous photodisintegration of infalling heavy nuclei from above and the launch of the $\nu_e$-burst associated with the shock propagation across the neutrinosphere of last scattering (see thick dashed green line in Fig.~\ref{fig:shellplot}). Consequently, the shock turns into an accretion front.

The supernova explosion, i.e. the revival of the shock wave and the subsequent ejection of the stellar mantle that surrounds the PNS, is due to the liberation of energy from the PNS interior to a thin layer of accumulated material at the PNS surface \citep[for recent reviews, cf.][]{Janka:2007,Janka:2012}. Several scenarios have been explored. Besides the magneto-rotational mechanism of \citet{LeBlanc:1970kg} \citep[for recent works, cf.][]{Takiwaki:2007sf,Winteler:2012,Moesta:2014,Moesta:2015} and the dumping of sound waves developed by \citet{Burrows:2005dv} -- yet not confirmed by other groups --  the neutrino-heating mechanism of \citet{Bethe:1985ux} has been demonstrated to lead to supernova explosions for a variety of massive progenitor stars \citep[cf.][]{Mueller:2012a,Takiwaki:2012,Bruenn:2013,Melson:2015,Lentz:2015,Roberts:2016}. However, in the framework of multi-dimensional simulations accurate Boltzmann neutrino transport cannot be employed due to the current computational limitations. Instead, approximate neutrino transport schemes are commonly used, whose range of applicability is currently being debated \citep[cf.][]{Sumiyoshi:2015}. Another issue of multi-dimensional supernova simulations with approximate neutrino transport may be related to the rather sparse neutrino phase-space resolution used, again, due to the current computational limitations. 

Besides the aforementioned three scenarios for the onset of the supernova explosion, another mechanism was discovered by \citet{Sagert:2008ka} due to the phase transition at high densities from ordinary nuclear matter to the quark gluon plasma. The latter was being treated within the simple but powerful thermodynamic bag model. During the phase transition a large amount of latent heat is released in a highly dynamical fashion, which in turn triggers the onset of the supernova explosion even in spherically symmetric simulations \citep[for details, see also][]{Fischer:2011}. Moreover, it leaves an observable millisecond burst in the neutrino signal \citep[for details, see][]{Dasgupta:2009yj}. These milestones demonstrate the sensitivity of EOS uncertainties related to our understanding of core-collapse supernovae and suffice the need of a more elaborate understanding of the EOS at high densities including better constraints in particular. 

Once the supernova explosion proceeds, mass accretion ceases at the PNS surface and the nascent PNS deleptonizes via the emission of neutrinos of all flavors on a timescale on the order of 10--30~s. This phase of the supernova evolution is mildly independent from details of the explosion mechanism. This has been explored in \citet{Fischer:2009af} and \citet{Huedepohl:2010} within the first self-consistent dynamical simulations based on accurate three-flavor Boltzmann neutrino transport in spherical symmetry. These aforementioned studies confirmed that the PNS settles into a quasi-static state~\citep[][]{Pons:1998mm} with a wind outflow developing from the PNS surface. This happens once the accretion funnels are quenched during the explosion phase and convection/SASI in the gain region cease.

 This is associated with the thick layer of low-density material accumulated at the PNS surface, which is subject to convection and dynamical modes before the supernova explosion onset, falling into the gravitational potential as mass accretion ceases. Also PNS convection affects the PNS deleptonization, which was studied in \citet{Roberts:2012f} and \citet{Mirizzi:2015eza}. Recently, it has been realized in sophisticated multi-dimensional long-term supernova simulations, with neutrino transport employed, that the beginning of the PNS deleptonization may be delayed by several seconds due to prevailing accretion flows onto the PNS surface \citep[cf.][]{Mueller:2015,Bruenn:2016}.

\subsection{Supernova EOS}
During the past years, constraints for model EOS for astrophysical applications have become increasingly stronger. Chiral effective field theory (EFT) \citep[cf.][and references therein]{Hebeler:2010a,Hebeler:2010b,Holt:2012a,Sammarruca:2012,Tews:2013,Krueger:2013,Coraggio:2013} is the ab-initio approach to the nuclear many-body problem of dilute neutron matter. It provides constraints up to normal nuclear matter density. Moreover, massive neutron stars with about 2~M$_\odot$ were observed  by \citet{Antoniadis:2013} and \citet{Demorest:2010}, recently reviewed by \citet{Fonseca:2016}, at high precision. Therefore, it requires sufficient stiffness at supersaturation densities. This finding challenges the appearance of additional particle degrees of freedom, e.g., hyperons and quarks. These tend to soften the EOS at supersaturation density. Note that this constraint ruled out the studies of \citet{Sagert:2008ka} and \citet{Fischer:2011} since their hadron-quark hybrid EOS yield maximum neutron star masses\footnote{~Here, hadron-quark hybrid stars, i.e. neutron stars with a quark core.} much below 2~M$_\odot$. On the other hand, the attempt of \citet{Fischer:2014} to construct a hybrid EOS based on the thermodynamic bag model in agreement with this constraint did not yield supernova explosions.

The large variety of conditions which are covered by the supernova EOS is illustrated in Fig.~\ref{fig:eos}. At temperatures below $\sim0.5$~MeV, time-dependent nuclear reactions determine the composition. There heavy nuclei dominate, being the ash from the advanced nuclear burning stages of the progenitor star. With increasing temperature, towards $T\simeq 0.5$~MeV, complete chemical and thermal equilibrium known as NSE (nuclear statistical equilibrium) is achieved. In NSE, the nuclear composition is determined from the three independent variables: temperature $T$, rest-mass density $\rho$ (or baryon number density, $n_{\rm B}$\footnote{~Restmass density $\rho$ and baryon number density $n_{\rm B}$ are related via $\rho=m_{\rm B} n_{\rm B}$, with $m_B$ being the baryon mass.}), and electron fraction $Y_e$. With increasing density, these nuclei become heavier while their abundance decreases simultaneously. At normal nuclear matter density, $\rho_0 \simeq 2.5\times10^{14}$~g~cm$^{-3}$ ($n_0\simeq0.15$~fm$^{-3}$), as well as above temperatures of about $5-10$~MeV, nuclei dissolve at the liquid-gas phase transition into homogeneous nuclear matter composed of quasi-free nucleons \citep[for details, cf.][]{Typel:2009sy,Hempel:2011,Roepke:2013}. 

The role of the EOS in core-collapse supernova simulations was explored in the multi-dimensional framework by \citet{Marek:2008qi}, \citet{Suwa:2013}, and recently by \citet{Nagakura:2017}, where neutrino-driven supernova explosions were the subjects of investigation. It was found that such explosions are favored for soft EOS, e.g., \citet{Lattimer:1991nc} with an earlier onset of shock revival and generally higher explosion energies, in comparison to stiff EOS, e.g., \citet{Shen:1998gg}. In failed core-collapse supernova explosions in spherical symmetry, EOS studies focused mainly on the dynamics and the neutrino signal up to the formation of the black hole \citep[cf.][]{Sumiyoshi:2006id,Fischer:2009,OConnor:2011,Steiner:2013}. Recently, the role of the nuclear symmetry energy in supernova simulations has been reviewed in \cite{Fischer:2014}. This is an important nuclear matter property that is recently becoming more tightly constrained by experiments, nuclear theory and observations \citep[for a summary of the current nuclear symmetry energy constraints, cf.][]{Lattimer:2013,Kolomeitsev:2016}.

\begin{figure}[btp!]
\subfigure[]{
\includegraphics[width=1\columnwidth]{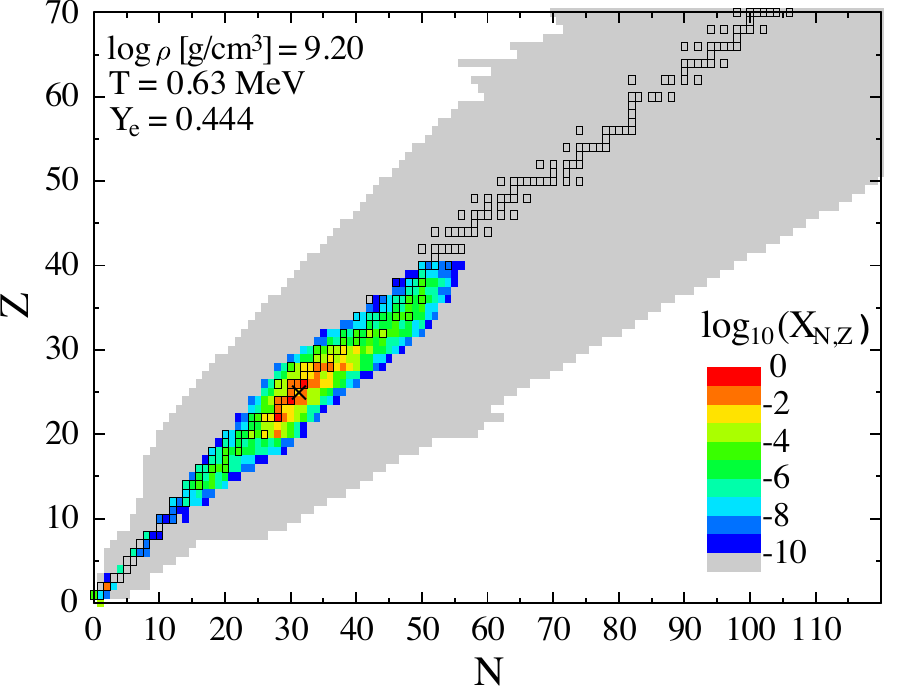}\label{fig:nse_a}}\\
\subfigure[]{
\includegraphics[width=1\columnwidth]{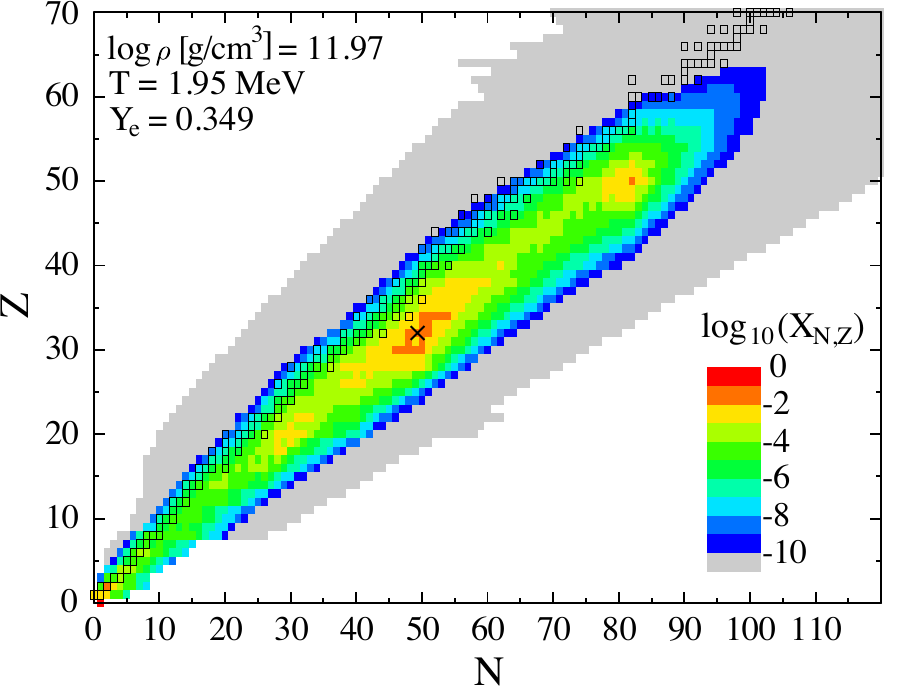}\label{fig:nse_b}}\\
\subfigure[]{
\includegraphics[width=1\columnwidth]{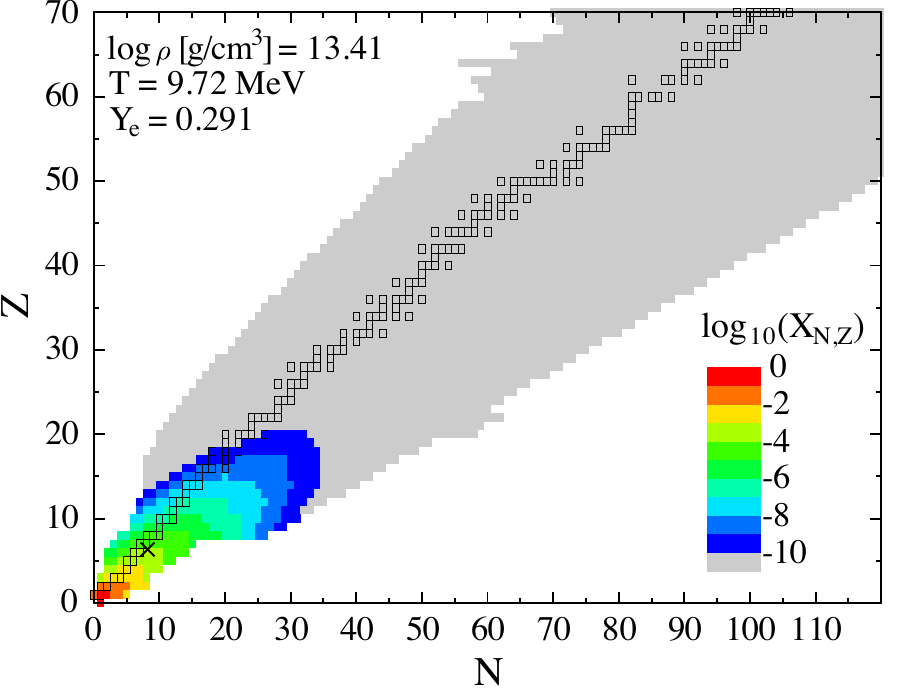}\label{fig:nse_d}}
\label{fig:neutrino}
\caption{Nuclear composition in the chart of nuclides (neutron number $N$ vs. proton number $Z$) based on the modified NSE approach of \citet{Hempel:2009mc}, obtained from the central conditions of the core-collapse evolution of \citet{Hempel:2012}.}
\label{fig:nse}
\end{figure}

\section{Heavy nuclei at low temperatures}
\label{sec2}
The domain of heavy nuclei can be sub-devided into two physically distinct conditions, i.e. where temperatures $T<0.5$~MeV -- time-dependent thermonuclear processes determine the nuclear composition -- and $T>0.5$~MeV where NSE is reached. 

\subsection{Small nuclear reaction networks}
In the regime of low densities and low temperatures ($T<0.5$~MeV), small nuclear reaction networks are commonly used which include about 14--50 nuclear species as explained in \citet{Thielemann:2004} \citep[the implementation of the network into our supernova model is discussed in][]{Fischer:2009af}. Even though they cannot accurately account for the evolution of $Y_e$ -- matter is nearly isospin symmetric with $Y_e\simeq 0.5$ (see the region below the horizontal dash-dotted line in Fig.~\ref{fig:eos}) -- they are sufficient for the nuclear energy generation. This domain, where time-dependent nuclear processes determine the evolution, corresponds to the outer core of the stellar progenitor, with the nuclear composition of dominantly silicon, sulphur as well as carbon and oxygen. In some cases, even parts of the hydrogen-rich helium envelope are taken into account, in particular in simulations of supernova explosions in order to be able to follow the shock evolution for tens of seconds through parts of the  stellar envelope. However, during the early post-bounce evolution prior to the supernova explosion onset, the stellar envelope remains nearly unaffected by the dynamics in the supernova core (see Fig.~\ref{fig:shellplot} above $10^3$~km). 

\subsection{NSE}
Towards the stellar core, the temperature increases above $T=0.5$~MeV where NSE is  fulfilled and where the relation $\mu_{(A,Z)}=Z\mu_p+(A-Z)\mu_n$ between the chemical potential of nucleus $\mu_{(A,Z)}$, with atomic mass $A$ and charge $Z$, and the chemical potentials of neutron $\mu_n$ and proton $\mu_p$ holds. The NSE conditions found in the collapsing stellar core feature a broad distribution of nuclei with a pronounced peak around the iron-group, at low densities (see Fig.~\ref{fig:nse_a}). These nuclei can be classified within the NSE average, including nuclear shell effects as discussed in \citet{Hempel:2009mc}. This method extends beyond the commonly used single-nucleus approximation, which is marked by the crosses in Fig.~\ref{fig:nse}. Note also that with increasing density the nuclear distribution shifts towards heavier nuclear species, moreover, it broadens with increasing temperature as illustrated in Fig.~\ref{fig:nse_b}. At high temperatures, around $T\simeq 5-10$~MeV, heavy nuclei dissolve via photodisintegration and (in)homogeneous nuclear matter forms, as shown in Fig.~\ref{fig:nse_d}. It will be discussed further in Sec.~\ref{sec3}.

\subsection{Weak interactions with heavy nuclei}
Heavy nuclei with nuclear charge $Z$ and mass $A$, are subject to fast electron captures, that are described collectively via the average composition as follows,
\begin{eqnarray}
e^-  + \langle A,Z \rangle \longrightarrow \langle A, Z-1 \rangle + \nu_e~.
\label{eq:e-capture}
\end{eqnarray}
This process deleptonizes stellar matter as the final-state neutrinos escape freely. As a consequence, the entropy per particle remains low during the entire stellar core collapse \citep[cf.][]{Bethe:1979,Lattimer:1981}, as illustrated in Fig.~\ref{fig:shellplot}. Electron capture rates commonly employed in supernova studies were developed by \citet{Bruenn:1985en} with a crudely simplified description of the Gamow window. Improved rates were provided by \citet{Juodagalvis:2010}, based on large-scale nuclear shell-model calculations including several 1000 nuclear species. A detailed comparison of both electron-capture rates and their impact on the collapse dynamics and neutrino signal can be found in \citet{Langanke:2003ii} and \citet{Hix:2003}. The rates of \citet{Juodagalvis:2010} are averaged over the NSE composition and provided to the community as a table.

In addition to electron captures, coherent neutrino-nucleus scattering is taken into account following \citet{Bruenn:1985en},
\begin{eqnarray}
\nu + \langle A,Z \rangle \leftrightarrows \langle A, Z \rangle + \nu~,
\label{eq:scatA}
\end{eqnarray}
for all flavors. This channel is essential for neutrino trapping once neutrinos are being produced with sufficiently high energies via~\eqref{eq:e-capture}. Inelastic neutrino-nucleus scattering rates were provided in \citet{Langanke:2007ua}. Moreover, heavy nuclei in the collapsing stellar core can exist at excited states, due to temperatures reached on the order of $0.5$~MeV up to few MeV. The subsequent nuclear de-excitation process via the emission of neutrino-antineutrino pairs,
\begin{eqnarray}
\langle A,Z \rangle^* \, \longrightarrow \, \langle A, Z \rangle + \nu + \bar\nu~,
\label{eq:dex}
\end{eqnarray}
can be undertsood in a similar fashion as the neutral-current process~\eqref{eq:scatA}.
The original idea, pointed out by \citet{Fuller:1991}, would potentially contribute to the losses during stellar collapse. In fact, in \citet{Fischer:2013} it was confirmed that process~\eqref{eq:dex} is the leading source of heavy lepton-flavor neutrinos as well as $\bar\nu_e$ during stellar core collapse. However, the neutrino fluxes remain small, compared to those of $\nu_e$ (see Fig.~\ref{fig:neutrino} top panel), and the neutrino energies are low, on the order of few MeV (see Fig.~\ref{fig:neutrino} bottom panel). Hence, the overall impact of nuclear de-excitations is negligible on the collapse dynamics and the neutrino signal. Instead, the stellar core collapse is dominated by nuclear electron captures and losses associated with $\nu_e$ (see Fig.~\ref{fig:neutrino}).

\begin{figure}[t!]
\includegraphics[width=1.0\columnwidth]{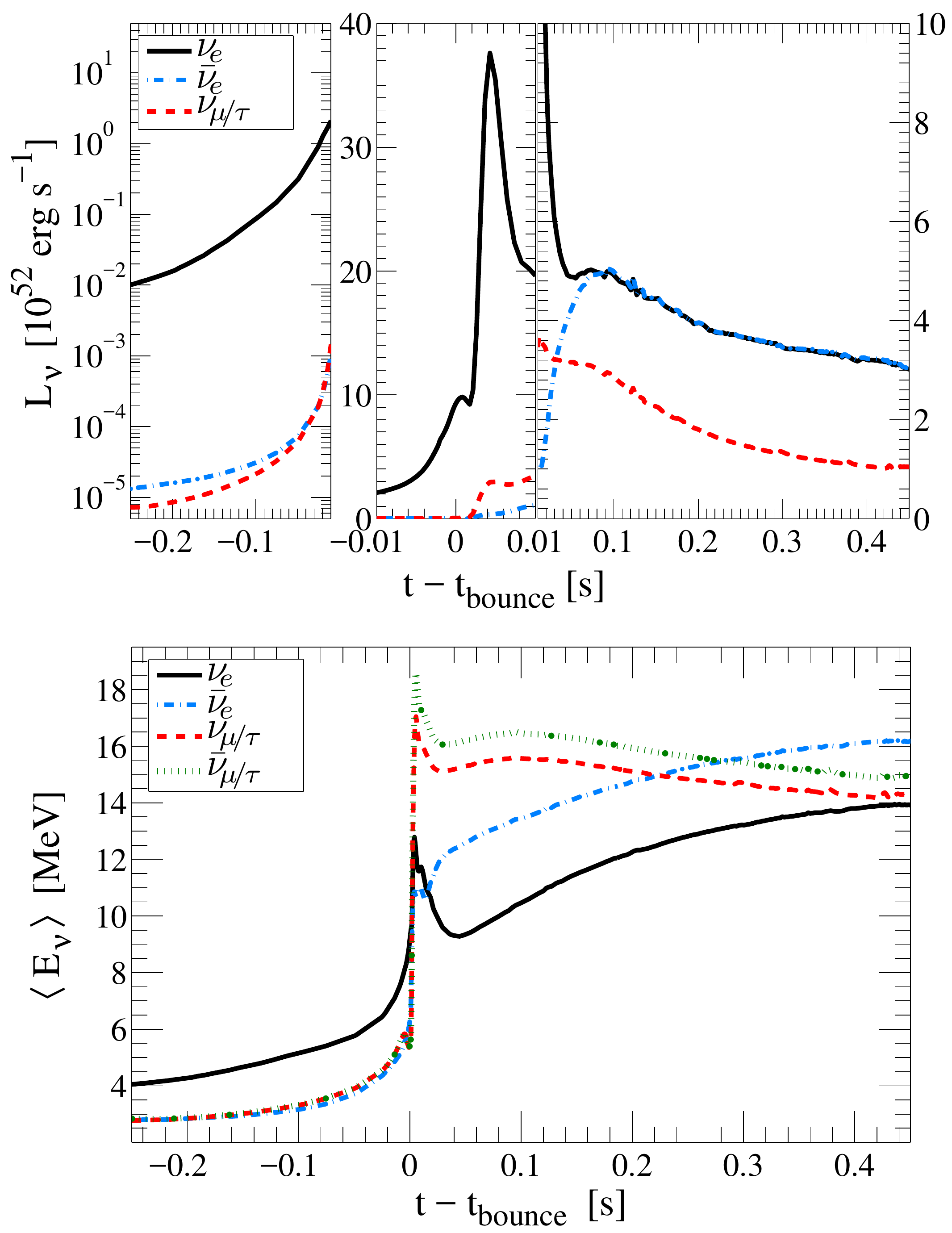}
\caption{Supernova neutrino signal, luminosities (top panel) and mean energies (bottom panel) for all flavors, sampled in the co-moving frame of reference at 1000~km. The supernova simulations were published in \cite{Fischer:2016a}, launched from the 18~M$_\odot$ progenitor of \citet{Woosley:2002zz}.}
\label{fig:neutrino}
\end{figure}

\subsection{Heavy nuclear structures at high density}
\begin{figure}[t!]
\centering
\includegraphics[width=1\columnwidth]{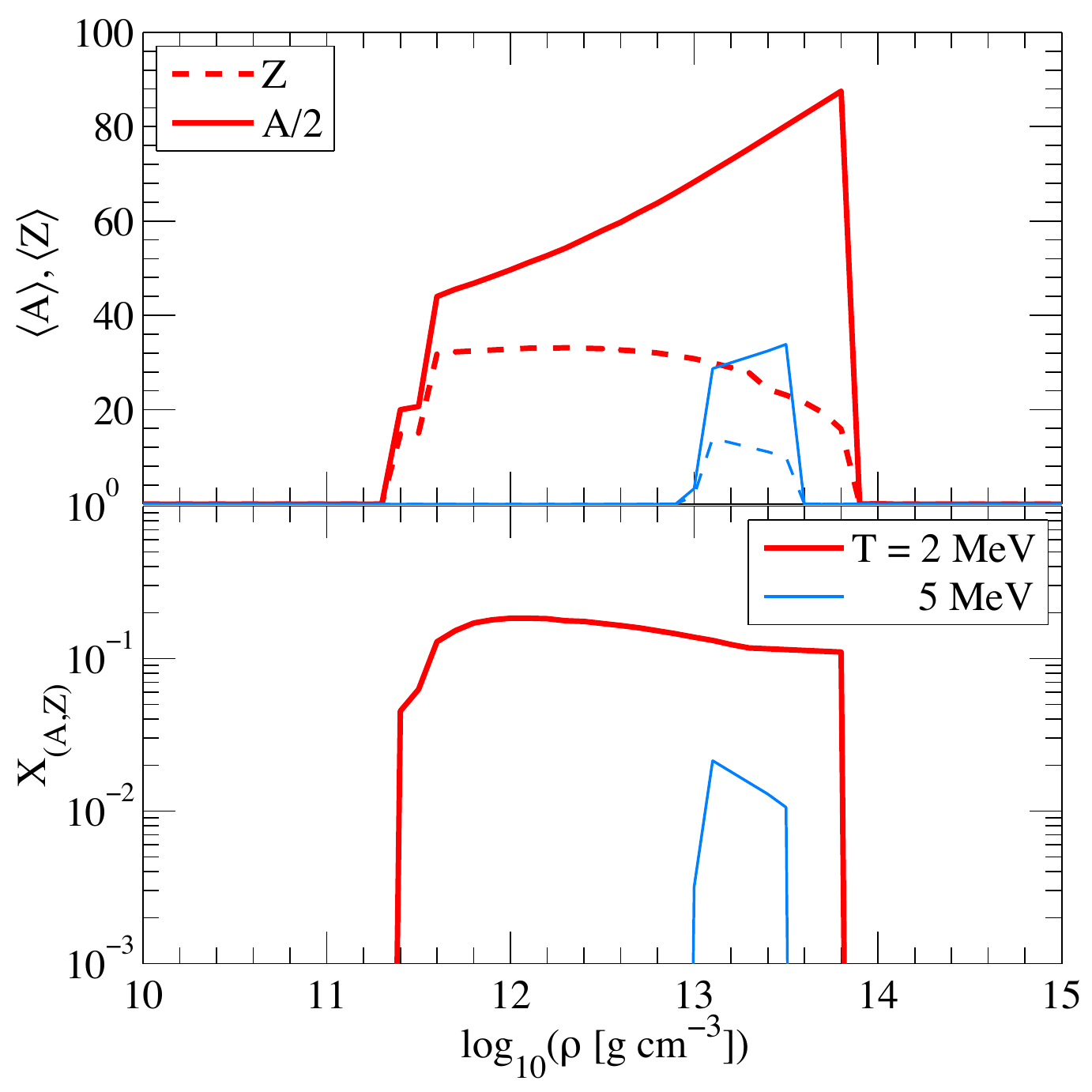}
\caption{Composition of heavy nuclear structures, average mass number as well as charge (top panel) and mass fraction, for matter in $\beta$-equilibrium at two selected temperatures, based on the Thomas-Fermi approximation of \citet{Shen:1998gg}. The increasing neutron excess visible is due to the continuously decreasing $Y_e$ with increasing density in $\beta$-equilibrium.}
\label{fig:pasta}
\end{figure}

With increasing density nuclei become heavier, as long as the temperatures are not too high which would enable efficient photodisintegration. This situation as well as the relevant density range is illustrated in Fig.~\ref{fig:pasta}, for matter in $\beta$-equilibrium at two selected temperatures, based on the Thomas-Fermi approximation of \citet{Shen:1998gg}. A detailed comparison between different nuclear approaches was performed in \citet{ShenG:2011}. Comparing the Thomas-Fermi approximation of \citet{Shen:1998gg}, compressible liquid drop model with Skyrme interactions by \citet{Lattimer:1991nc} and the virial EOS with nucleons and nuclei of \citet{ShenG:2010b} \citep[combined with the relativistic mean field EOS of][]{ShenG:2010a}, qualitative agreement was found for the gross properties, e.g., pressure, entropy, nuclear abundances, average nuclear mass and charge, comparing the three models. 

Note that at the density range where these nuclear structures appear (see Fig.~\ref{fig:pasta}), all protons in the system are consumed into heavy nuclei, such that effectively only free neutrons exist besides heavy and light nuclei. The latter aspect will be further discussed below in Sec.~\ref{sec4}. The situation illustrated in Fig.~\ref{fig:pasta} corresponds to the liquid-gas phase transition at finite temperatures and large isospin asymmetry (note the very low proton abundances of $Y_p=0.01-0.1$ in $\beta$-equilibrium in this density domain).

It has long been realized that the formation of the heavy nuclear structures sketched in Fig.~\ref{fig:pasta} via spherical heavy nuclei, are due to the competition of the attractive long-range nuclear force and Coulomb repulsion \citep[cf.][and references therein]{Watanabe:2003,Watanabe:2009,Newton:2009,Molinelli:2014a,Horowitz:2014}. Due to surface effects, these structures form shapes, e.g., spaghetti, lasagna and meat balls, denoted collectively as ''nuclear pasta''. The conditions where nuclear pasta phases exist are listed in Table~\ref{tab:pasta} at two selected values of $Y_e$, based on the detailed 3-dimensional Skyrme-Hartree-Fock calculations of \citet{Newton:2009}. Comparing the density ranges of Table~\ref{tab:pasta} for $Y_e=0.05$ and Fig.~\ref{fig:pasta}, it becomes clear that only towards high density nuclear pasta appears, where at low densities spherical heavy nuclei exist. The reason why the heavy structures dissolve already below $\rho=10^{14}$~g~cm$^{-3}$ in Fig.~\ref{fig:pasta} is due to the even lower $Y_e$ in $\beta$-equilibrium, which is assumed in Fig.~\ref{fig:pasta}. This points to the very sensitive dependence of nuclear pasta phases on temperature and $Y_e$.  Moreover, it has been realized that the neutrino mean free path is modified in nuclear pasta. Detailed molecular dynamics simulations of the neutrino response from coherent neutrino scattering were conduced in \citet{Horowitz:2004a} and \citet{Horowitz:2004b}. An alternative approach has been developed in \citet{Molinelli:2014b}.

\begin{table}[t!]
\centering
\begin{tabular}{c c c}
\hline
\hline
$Y_e$ & Density range & $T_{\rm melt}$ \\
& $[\rm g~\rm gm^{-3}]$ & $[\rm MeV]$ \\
\hline
0.05 & $\sim 8.3\times 10^{13} - 1.3 \times 10^{14}$ & $\sim 3-5$ \\
0.30 & $\sim 3.3\times 10^{13} - 2.0 \times 10^{14}$ & $\sim 10$ \\
\hline
\end{tabular}
\caption{Selected conditions for the presence of nuclear pasta, in terms of two values of $Y_e$ and the density range, from calculations based on \citet{Newton:2009}. $T_{\rm melt}$ marks the approximate melting temperatures.}
\label{tab:pasta}
\end{table}

Note that this phase is relevant for the post-bounce supernova evolution prior to the explosion onset, since temperatures in this density range exceed $T=5$~MeV and hence pasta melts (see Fig.~\ref{fig:pasta}). However, the situation changes during the later PNS deleptonization, after the supernova explosion onset when the temperature decreases continuously. Even though the structure of the PNS is not affected by the presence of such heavy nuclear structures, neutrino interactions may well modify the timescale on which neutrinos diffuse out of the PNS interior. Therefore, the very first detailed supernova simulations with sophisticated neutrino transport and an effective description of coherent neutrino-pasta scattering have been presented recently in \citet{Horowitz:2016}. These results show qualitatively the role of nuclear pasta, i.e. an extended deleptonization and cooling time of the PNS, once pasta phases form.

\section{Inhomogeneous nuclear matter}
\label{sec3}
During the core collapse evolution temperature and density rise continuously, which eventually leads to the transition to inhomogeneous matter with light and heavy nuclear clusters (corresponding to the transition from Fig.~\ref{fig:nse_b} to Fig.~\ref{fig:nse_d}). The conditions for this transition are obtained already before core bounce, and they remain during the entire post-bounce evolution, located between the supernova shock  and the PNS surface (see the region of high entropy in Fig.~\ref{fig:shellplot}). This corresponds to the conditions where neutrinos decouple from matter and hence a 'good' treatment of weak processes and nuclear medium is essential. Weak reactions with heavy nuclei play only a sub-dominant role. Heavy nuclei dissociate due to the high temperatures and weak processes with free nucleons are significantly faster.

\subsection{Weak processes}
Here we distinguish electronic charged-current processes,
\begin{eqnarray}
e^- + p\,\leftrightarrows\, n + \nu_e~,\;\;\;e^+ + n \, \leftrightarrows \, p + \bar\nu_e~,
\label{eq:cc}
\end{eqnarray}
neutral-current elastic scattering on nucleons ($N$),
\begin{eqnarray}
\nu + N \, \leftrightarrows \, N + \nu~,
\label{eq:scat1}
\end{eqnarray}
inelastic scattering on electrons and positrons,
\begin{eqnarray}
\nu + e^\pm  \leftrightarrows \, e^\pm + \nu~,
\label{eq:scat2}
\end{eqnarray}
and pair processes,
\begin{subequations}
\begin{eqnarray}
e^- + e^+ \leftrightarrows \nu + \bar\nu~,\;\;
N N \leftrightarrows N N \, \nu + \bar\nu~,
\label{eq:pair}
\end{eqnarray}
\begin{eqnarray}
\nu_e + \bar\nu_e \leftrightarrows \nu_{\mu/\tau} + \bar\nu_{\mu/\tau}~,
\label{eq:pair_nu}
\end{eqnarray}

\end{subequations}
where $\nu\in\{\nu_e,\bar\nu_e,\nu_{\mu/\tau},\bar\nu_{\mu/\tau}\}$ and $N\in\{n,p\}$ else notified otherwise. In \citet{Buras:2002wt} additional inelastic scattering processes have been considered, in analogy to the process \eqref{eq:pair_nu},
\begin{eqnarray}
\nu_{\mu/\tau} + \nu_e \, \leftrightarrows \, \nu_e + \nu_{\mu/\tau}~,\,\,\,
\bar\nu_{\mu/\tau} + \nu_e \, \leftrightarrows \, \nu_e + \bar\nu_{\mu/\tau}~, \\
\nu_{\mu/\tau} + \bar\nu_e \, \leftrightarrows \, \bar\nu_e + \nu_{\mu/\tau}~,\,\,\,
\bar\nu_{\mu/\tau} + \bar\nu_e \, \leftrightarrows \, \bar\nu_e + \bar\nu_{\mu/\tau}~,
\label{eq:scat3}
\end{eqnarray}
which are relevant at high densities and temperatures where a large trapped $\nu_e$ component exists. Recently, in \citet{Fischer:2016b} the inverse neutron decay has been implemented,
\begin{eqnarray}
n \, \leftrightarrows \, p + e^- + \bar\nu_e~.
\label{eq:ndec}
\end{eqnarray}
Reactions~\eqref{eq:cc} and \eqref{eq:ndec} are known as {\em Urca} processes. Together with \eqref{eq:scat1}, they are typically treated in supernova simulations within the zero-momentum transfer approximation of \citet{Bruenn:1985en}. Inelastic contributions as well as corrections from weak magnetism are taken into account effectively in present supernova studies, following \citet{Horowitz:2001xf}, which also takes into account the strangeness contents in the baryons via a strangeness axial-vector coupling constant, $g_S$, which effectively reduces the axial-vector coupling constant, $g_A-g_S$. The currently accepted value for the nucleon strangeness contents, deduced from deep-inelastic proton-scattering experiments, relates to values of $g_S\simeq 0.1$ \citep[cf.][and references therein]{Hobbs:2016}. Note that weak magnetism enhances the opacity for $\nu$ while it suppresses the opacity for $\bar\nu$. It leads to the non-negligible enhancement of spectral differences between $\nu$ and $\bar\nu$, in particular for the heavy lepton flavor neutrinos where it is the leading cause, besides neutrino electron/positron scattering \eqref{eq:scat2}. The pair processes \eqref{eq:pair} and \eqref{eq:pair_nu} don't distinguish between $\nu$ and $\bar\nu$, i.e. both are produced with identical spectra. 

\begin{figure}[t!]
\includegraphics[width=1\columnwidth]{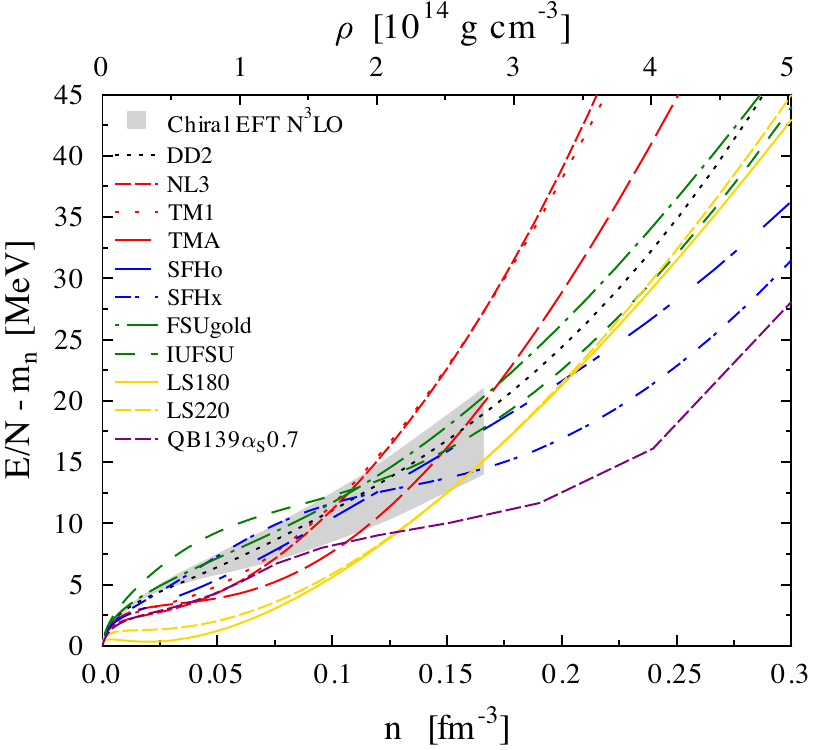}
\caption{Neutron matter energy per particle for a selection of supernova model EOS, in comparison to the chiral EFT constraint of \citet{Krueger:2013}. See text for details. \citep[Figure adopted from][]{Fischer:2014}}
\label{fig:enm}
\end{figure}

For the processes~\eqref{eq:cc} and \eqref{eq:ndec} it is important to treat these weak interactions consistently with the underlying nuclear EOS, which was pointed out by \citet{MartinezPinedo:2012} and \citet{Roberts:2012} based on the mean-field description of \citet{Reddy:1998}. The associated medium modification, $\triangle U = U_n - U_p$, defines the difference between neutron and proton single particle potentials \citep[i.e. vector self energies within the relativistic mean field (RMF) framework as was discussed in, e.g.,][]{Hempel:2015b}. They depend on the nuclear symmetry energy $\triangle U \propto E_{\rm sym}(T,\rho)$ which has a strong density dependence. A detailed comparison between neutron matter and symmetric matter EOS can be found in \citet{Typel:2014b}. Moreover, it was confirmed in detailed supernova simulations that $E_{\rm sym}$ determines the spectral difference between $\nu_e$ and $\bar\nu_e$, however, with relevance only during the PNS deleptonization after the supernova explosion onset has been launched. This is related to the energy scales involved. Note the $Q$-values for processes~\eqref{eq:cc}: $Q=\pm Q_0\pm \triangle U$, for $\nu_e (+)$ and $\bar\nu_e(-)$. $Q_0$ denotes the vacuum $Q$-value of the processes, $Q_0=m_n-m_p=1.2935$~MeV, being  the neutron-to-proton restmass difference. At low densities, the energetics of the processes~\eqref{eq:cc} is determined by $Q_0$, since $\triangle U \ll Q_0$. With increasing density the medium modifications start to dominate when $\triangle U \gtrsim Q_0$.

In order to determine $(U_n,U_p)$ it is essential for supernova simulations to employ model EOS with a ''good'' low-density behavior. Here, the ab initio approach is chiral~EFT of dilute neutron matter. Fig.~\ref{fig:enm} illustrates the chiral~EFT results from \citet{Krueger:2013} together with a selection of RMF model EOS (DD2--IUFSU) and the non-relativistic EOS (LS180 and LS220), which were and still are commonly used in supernova simulations. Details about these EOS and a table that lists a selection of nuclear matter properties can be found in \citet{Fischer:2014}. Important here is the maximum neutron star mass constraint of $2$~M$_\odot$ (not fulfilled by FSUgold, IUFSU and LS180) and the constraint of the nuclear symmetry energy and its slope parameter at nuclear saturation density \citep[fulfilled by DD2 and SFHo, for details cf.][]{Lattimer:2013}. Moreover, from Fig.~\ref{fig:enm} it becomes evident that the RMF model with density-dependent couplings DD2 of \citet{Typel:2009sy} is in quantitative agreement with chiral~EFT. The other two EOS in good agreement with chiral~EFT are the EOS of \citet{Steiner:2013}, SFHo and SFHx, which were developed in accordance with neutron star radii deduced from the analysis of low-mass X-ray binaries by \citet{Steiner:2010}. All other EOS, including the quark matter EOS of \citet{Fischer:2014} based on the thermodynamic bag model (QB139$\alpha_S$0.7 -- we will come back to quark matter EOS in more details in Sec.~\ref{sec5}), violate the chiral~EFT constraint, besides the aforementioned conflicts with the other constraints. 

In addition to the mean-field effects, which modify the charged-current processes, nuclear many-body correlations suppress the charged-current absorption rates and neutral-current neutrino scattering processes~\eqref{eq:scat1} with increasing density, for which the expressions of \citet{Burrows:1998} and \citet{Burrows:1999} are commonly employed in supernova studies. Recently, \citet{Horowitz:2017} reviewed many-body correlations for the neutral-current neutrino-nucleon scattering processes. The authors provide a useful fit for the vector response function, in combination with the Random Phase Approximation at high densities \citep[see also][and references therein]{Reddy:1999, Roberts:2017} and the virial EOS for the low-density part.

Reaction rates for pair processes are provided in \cite{Bruenn:1985en}, where as $N$--$N$-bremsstrahlung rates were developed in \citet{Hannestad:1997gc} based on the vacuum $1\pi$-exchange framework developed by \citet{Friman:1979}. Recently, \citet{Fischer:2016b} extended this treatment of the vacuum $1\pi$-exchange for $N$--$N$-bremsstrahlung by taking into account the leading-order medium modifications, i.e. dressing of the $\pi NN$-vertex. Based on the Fermi-liquid theory. Expressions have been derived that can be implemented into supernova simulations. In analogy, \citet{Bartl:2016} describe such medium modifications at the level of chiral~EFT. 

The annihilation of trapped $\nu_e\bar\nu_e$ pairs processes of \eqref{eq:pair_nu} couples electron and heavy lepton flavor neutrinos, at high temperatures and densities.  This channel reduces the difference of the luminosities and average energies of both flavors. Reaction rates were implemented in \citet{Buras:2002wt} and \citet{Fischer:2009}. Moreover, the highly inelastic neutrino-electron(positron) scattering \eqref{eq:scat2} thermalizes the neutrino spectra. Expressions for weak rates can be found, e.g., in \citet{Mezzacappa:1993gx}.

\begin{figure}[t!]
\includegraphics[width=1\columnwidth]{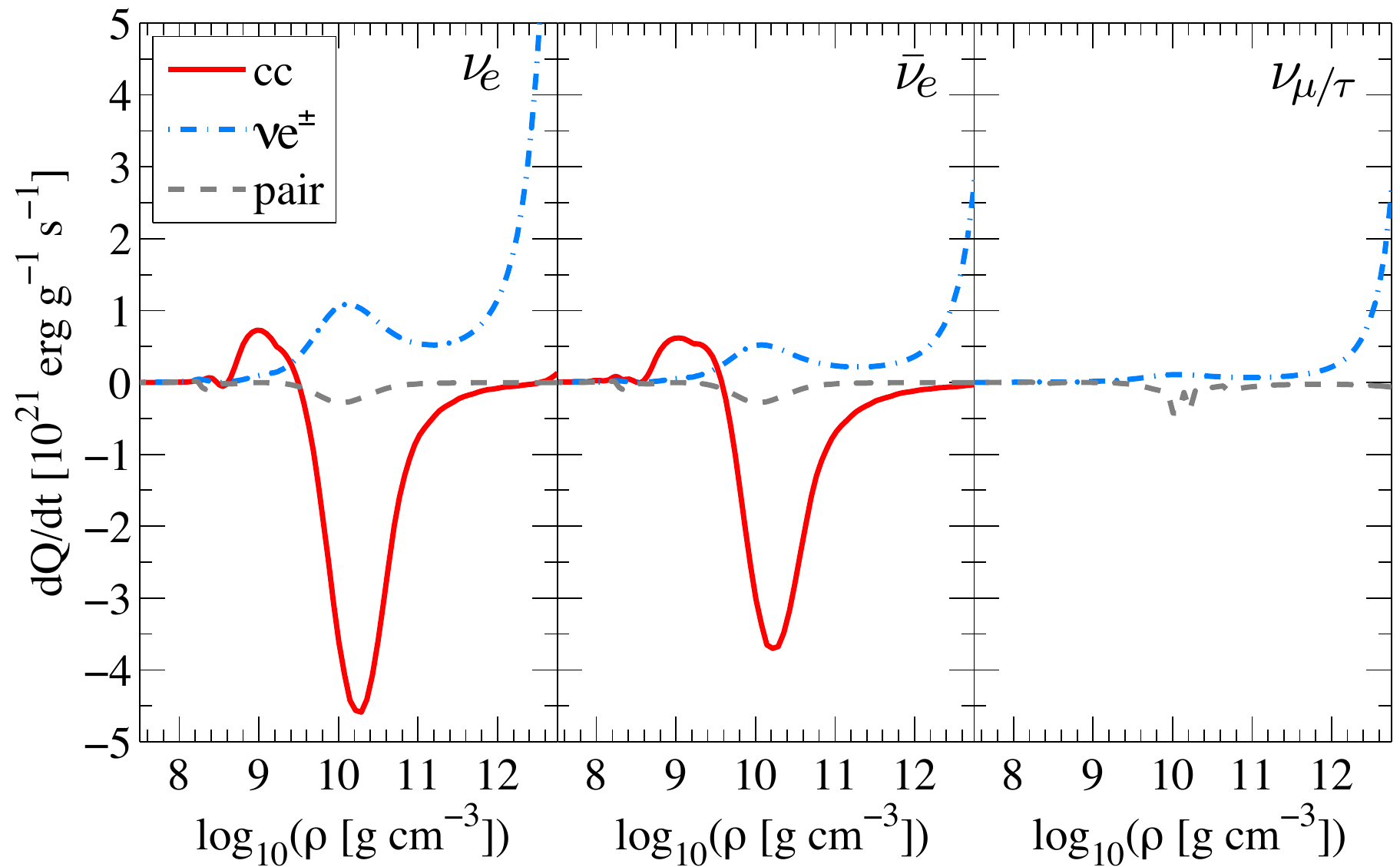}
\caption{Integrated neutrino heating ($dQ/dt>0$) and cooling ($dQ/dt<0$) rates of the different channels charged-current (cc) processes~\eqref{eq:cc}, neutrino-electron and positron scattering ($\nu e^\pm$) processes~\eqref{eq:scat2}, and the sum of all pair reactions (pair) processes~\eqref{eq:pair}. The data are from the reference supernova simulation of \citet{Fischer:2016a} as illustrated in Fig.~\ref{fig:shellplot} at about 300~ms post bounce, and the density domain corresponds to the region between PNS surface at around 15--20~km and the standing bounce shock around 80~km.}
\label{fig:heating}
\end{figure}

\subsection{Post-bounce supernova dynamics and neutrino emission}
Reaction~\eqref{eq:cc} is responsible for the launch of the $\nu_e$-burst (top central panel in Fig.~\ref{fig:neutrino}), which is associated with the propagation of the bounce shock across the neutrinospheres of last scattering (see Fig.~\ref{fig:shellplot}) between 5--20~ms after core bounce. Weak equilibrium is re-established as matter is shock heated, associated with the sudden rise of the temperature. This highly non-equilibrium phenomenon is essential for the following supernova evolution as it determines a major source of losses, several $10^{53}$~erg~s$^{-1}$, being partly responsible for the dynamic bounce shock turning into a standing accretion front (see Fig.~\ref{fig:shellplot}). Moreover, only slightly before core bounce, when positrons exist, all other neutrino flavors are being produced mainly via electron-positron annihilation  as well as via nucleon-nucleon bremsstrahlung (pair processes~\eqref{eq:pair}). The luminosities of $\bar\nu_e$ and heavy- lepton flavor neutrinos rise accordingly (see Fig.~\ref{fig:neutrino}). The luminosities of all heavy lepton flavor neutrinos rise somewhat faster than those of $\bar\nu_e$, since the latter are coupled more strongly to matter via the charged-current channel (see the second process of \eqref{eq:cc}). This feature will eventually allow us to probe the neutrino mass hierarchy via the neutrino signal rise time from the neutrino observation of the next galactic supernova event \citep[details can be found in][]{Serpico:2011ir}.

The later post-bounce evolution is determined by mass accretion onto the standing shock (see Fig.~\ref{fig:shellplot}), during which the average neutrino energy hierarchy is determined by the different coupling strengths to matter (see bottom panel in Fig.~\ref{fig:neutrino}). Consequently, each neutrino species has their own neutrinosphere radius $R_{\nu}$ of last (in)elastic collision where the following hierarchy holds: $R_{\nu_e}>R_{\bar\nu_e}>R_{\nu_{\mu/\tau}}\gtrsim R_{\bar\nu_{\mu/\tau}}$. The electron (anti)neutrinos decouple in a thick layer of low-density material accumulated at the PNS surface, powered by the charged-current processes~\eqref{eq:cc}. Consequently, their luminosity can be approximated by the accretion luminosity as follows \citep[cf.][]{Janka:2007,Janka:2012},
\begin{eqnarray}
L_{\nu_e} \propto 10^{52}
\left(\frac{{M}}{1.5~\rm M_{\odot}} \right)
\left(\frac{\dot{m}}{0.4~\rm \frac{M_{\odot}}{{\rm s}}} \right)
\left(\frac{100~\rm km}{R_{\nu_e}}\right)
\frac{\rm erg}{\rm s} ~,
\end{eqnarray}
with a typical mass enclosed inside the PNS $M$ and radius associated with the neutrinospehere $R_{\nu_e}$ as well as mass accretion rate $\dot m$. On the other hand, the heavy lepton neutrino flavors are determined by diffusion in the absence of charged-current processes. 

The process of neutrino decoupling from matter is neutrino transport problem. Accurate three-flavor Boltzmann neutrino transport has been developed for spherically symmetric supernova models in \citet{Mezzacappa:1993gn} and \citet{Liebendoerfer:2004}. It leads to the establishment of a large cooling layer towards high densities at the PNS surface, where $dQ/dt<0$\footnote{~We give a brief description of the cooling/heating rates in the Appendix~\ref{appendix}.}, as illustrated in Fig.~\ref{fig:heating}. It corresponds to the domain where high energy neutrinos decouple from matter, while the low energy spectrum is still thermalized with the medium. At low densities, between the standing shock at around $10^9$~g~cm$^{-3}$, these low energy neutrinos deposit parts of their energy via absorption processes into the medium. There, a heating layer establishes where $dQ/dt>0$. However, since most weak interaction rates have a strong dependence on the neutrino energy, the integrated heating rates are significantly smaller than the cooling at higher density, besides the smaller mass enclosed in the heating layer than in the cooling layer. These are the two main reasons why spherically symmetric supernova explosions could not be obtained for the massive progenitor stars that develop an extended mass accretion period, typically for stars with initial mass above around 10~M$_\odot$. The success of the neutrino-heating mechanism in multi-dimensional simulations is attributed to the development of convection which allows material to remain effectively longer in the heating region, which increased the neutrino-heating efficiency. However, it should be mentioned that up to now only neutrino transport approximation schemes have been employed in multi-dimensional supernova studies. Note further that the situation is different for the low progenitor mass range, between $8-10$~M$_\odot$. Such stars develop either oxygen-neon cores \citep[cf.][]{Nomoto:1987,Jones:2013}, leading to electron-capture supernovae as explored in \citet{Kitaura:2006} and \citet{Fischer:2009af}, or ''tiny'' iron-cores as was explored recently in \citet{Melson:2015}. In both cases, the special structure of the stellar core, i.e. sharp density gradient separating core and envelope, leads to fast shock expansions and explosions even in spherical symmetric supernova simulations with low explosion energies $\sim 10^{50}$~erg and small amount of nickel ejected \citep[for details, see][]{Wanajo:2009}. Similar core structures are obtained from binary stellar evolution. This was explored in \citet{Tauris:2013} and \citet{Tauris:2015}, leading to so-called ultra-stripped progenitors of the secondary star that has undergone major mass transfer during the common-envelope evolution.

\section{Role of light nuclear clusters}
\label{sec4}
With the recent advance regarding the improved description of medium modified nuclei \citep[cf.][]{Roepke:2013}, particular interest has been devoted to the question about light nuclear clusters \citep[see also][for a recent discussion about light clusters in heavy-ion collision experiments as tracers of early flow]{Bastian:2016}. It concerns nuclei with mass numbers $A=2-4$. About their role on the supernova dynamics and the neutrino signal has long been speculated. 

\begin{figure}[t!]
\centering
\includegraphics[width=1.0\columnwidth]{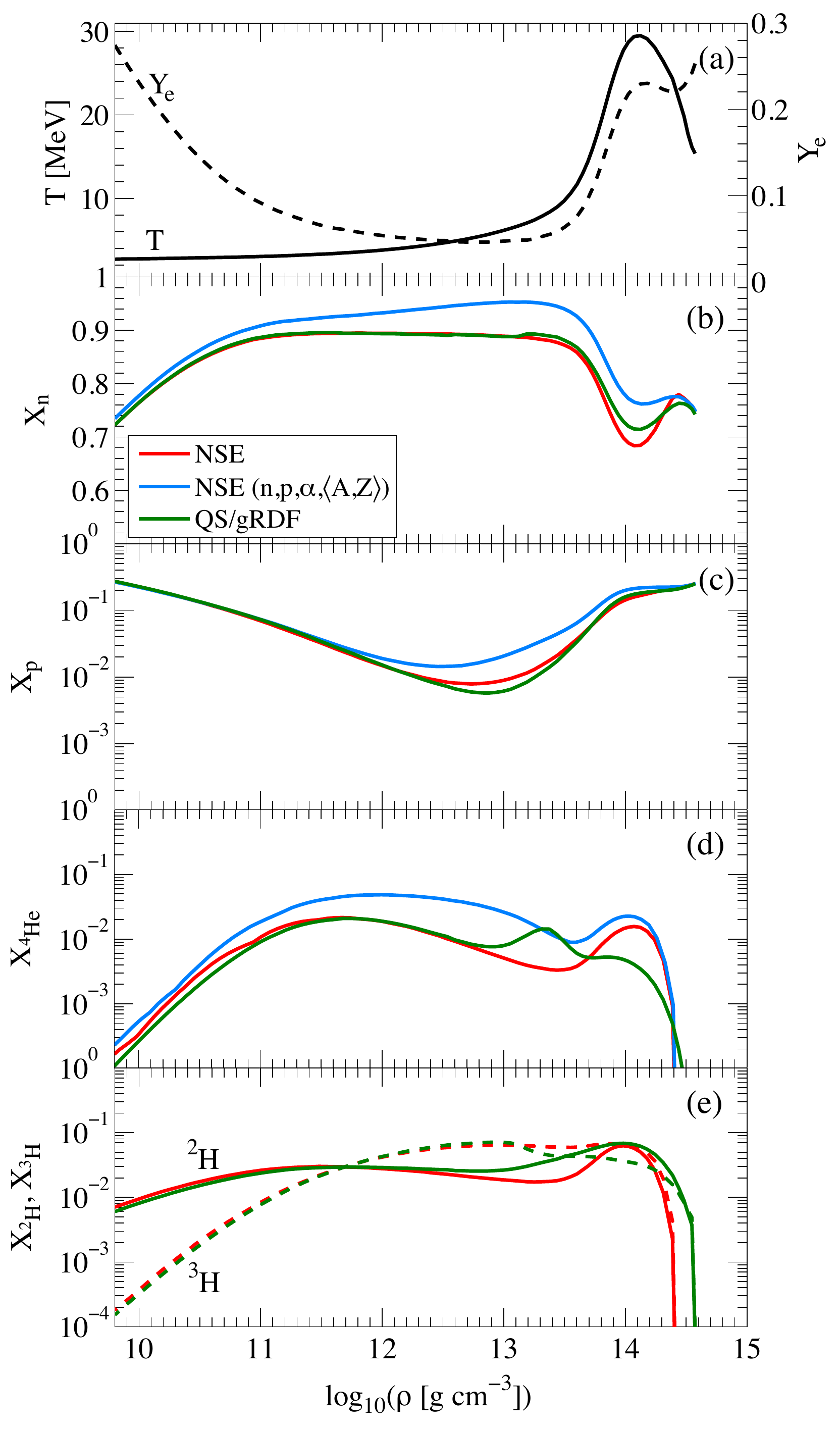}
\caption{Abundance of the standard supernova composition(middle panel) in comparison to the light clusters deuteron and triton (bottom panel). The conditions correspond to the early PNS deleptonization phase shortly after the supernova explosion onset has been launched (top panel) when the abundance of light nuclei with $A=2-3$ is maximum relative to those of protons \citep[data obtained from][]{Fischer:2016d}.}
\label{fig:light_abn}
\end{figure}
\begin{figure}[t!]
\centering
\includegraphics[width=1.0\columnwidth]{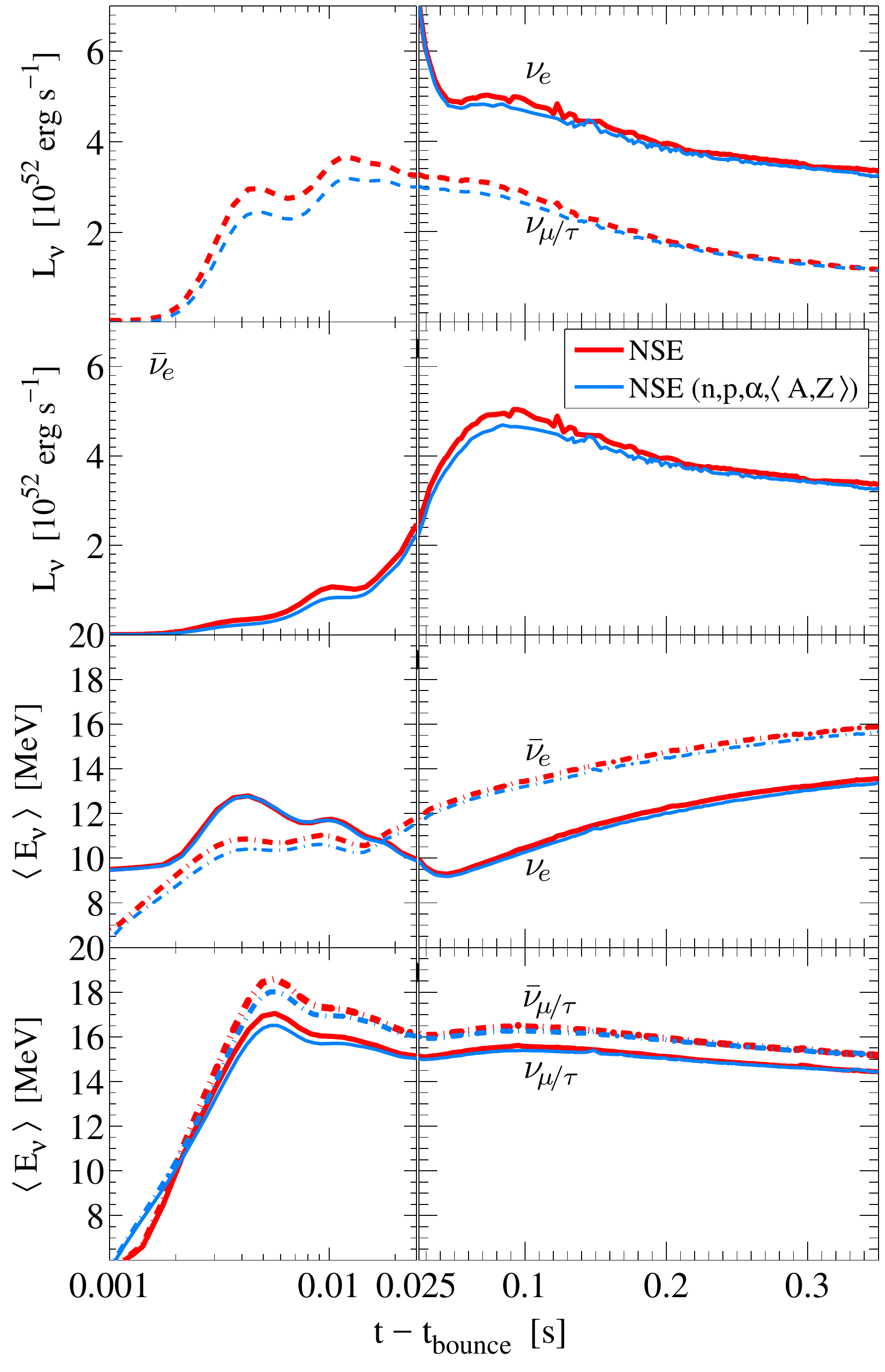}
\caption{Neutrino luminosities and average energies sampled in the co-moving frame of reference at 1000~km, comparing simulations where 'all' nuclear clusters are included based on the modified NSE approach of \citet{Hempel:2009mc} (same as shown in Fig.~\ref{fig:neutrino}) with the simplified composition $(n,p,\alpha,\langle A,Z \rangle)$.}
\label{fig:neutrino_npaA}
\end{figure}

Within the 'classical' nuclear setup for supernova EOS, e.g., based on \citet{Lattimer:1991nc} and \citet{Shen:1998gg}, the simplified nuclear composition includes free nucleons, $\alpha$ particle and a single representative heavy nucleus with average mass and charge, $(n,p,\alpha,\langle A,Z \rangle)$. Hence, the question about the role of light clusters, others than $\alpha$ particles, could not be attributed. A first attempt to include all light clusters was given by \citet{Sumiyoshi:2008} where the quantum statistical approach of \citet{Roepke:1982} has been used. Based on the concept of the excluded volume, already used in the 'classical' EOS, an advanced supernova EOS was developed by \citet{Hempel:2009mc} with the inclusion of a detailed nuclear composition. The conditions where light clusters with $A=2-4$ are abundant corresponds to the region of high entropy between the neutrinospheres and the bounce shock, see Fig.~\ref{fig:shellplot}, denoted here collectively via $^4$He.

There are two crucial aspects related to light clusters: {\em (a)} modification of the nuclear EOS and {\em (b)} the inclusion of a large variety of weak processes \citep[cf. right column of Table~(1) in][]{Fischer:2016d} in addition to the standard weak processes~\eqref{eq:e-capture}--\eqref{eq:ndec}. 

\subsection{EOS with light clusters}
The consistent description of the nuclear medium with light clusters as explicit degrees of freedom reduces the abundance of the free nucleons and $^4$He, in the domain where light clusters are abundant. This is illustrated in Fig~\ref{fig:light_abn} (middle panel) at selected conditions found during the early PNS deleptonization shortly after the supernova explosion onset -- temperature and $Y_e$ profiles are shown in the top panel, with respect to the baryon density. Here, we compare the medium-modified NSE approach of \citet{Hempel:2009mc} including the complete abundances of all nuclear clusters (red lines) with those of the same approach where only $^4$He is considered as light nuclear cluster (blue lines). Note that the latter case corresponds to the 'classical' supernova EOS composition that was commonly employed in numerous supernova simulations. It becomes evident that not only the abundance of $^4$He is largely overestimated, also the abundances of neutrons and protons are overestimated. Note further that the region where light clusters and free protons are equally abundant, corresponds to the supernova cooling region (see Fig.~\ref{fig:heating}). Hence, this urges the need for the systematic comparison of EOS based on the 'full' composition and only simplified nuclear composition $(n,p,\alpha,\langle A,Z \rangle)$, in supernova simulations in much greater detail, especially within multi-dimensional framework. In particular, we find that weak reactions with protons, most relevant for the charged-current $\bar\nu_e$-opacity, are off by a factor greater than two, since the reaction rates scale with the number density of protons. The magnitude of the differences in spherically symmetric supernova simulations is illustrated in Fig.~\ref{fig:neutrino_npaA}, where we compare the modified NSE approach of \citet{Hempel:2009mc} with 'all' nuclear clusters included with the simplified composition $(n,p,\alpha,\langle A,Z \rangle)$. In particular the luminosity and average energy of $\bar\nu_e$ are overestimated when considering the simplified composition. Moreover, the rise time of the neutrino signal \citep[for details about the role of the neutrino rise time can be found in][]{Serpico:2011ir}, in particular for $\bar\nu_e$ and heavy-lepton flavor neutrinos, is suppressed with $(n,p,\alpha,\langle A,Z \rangle)$, being related to the suppression of $N$--$N$-bremsstrahlung processes. This may have implications for the appearance of prompt convection, which occurs on a short timescale on the order of few tens of milliseconds after core bounce. The potential impact remains to be explored in multi-dimensional simulations.

In Fig.~\ref{fig:light_abn} we also compare the modified NSE of \citet{Hempel:2009mc} with the more sophisticated approaches for the description of in-medium nuclear clusters, i.e. the generalized RMF approach (gRDF) of \citet{Typel:2009sy}. The latter is based on in-medium nuclear properties, e.g., binding energies obtained within first-principle quantum statistical calculations of \citet{Roepke:2009} and \citet{Roepke:2011}. There it becomes evident that the modified NSE approach provides a sufficient description of the growth properties, such as the particle densities, of light nuclear clusters as illustrated in Fig.~\ref{fig:light_abn}~(c). However, the caveat is at high densities, $\rho>10^{14}$~g~cm$^{-3}$, where the geometric excluded volume of the modified NSE fails to properly describe the dissolving of nuclear states into the mean field. Modified NSE provides an inaccurate description of the phase transition to homogeneous matter with over- and underestimated abundances of the light clusters, depending on density and temperature.

\begin{figure}[t!]
\includegraphics[width=1\columnwidth]{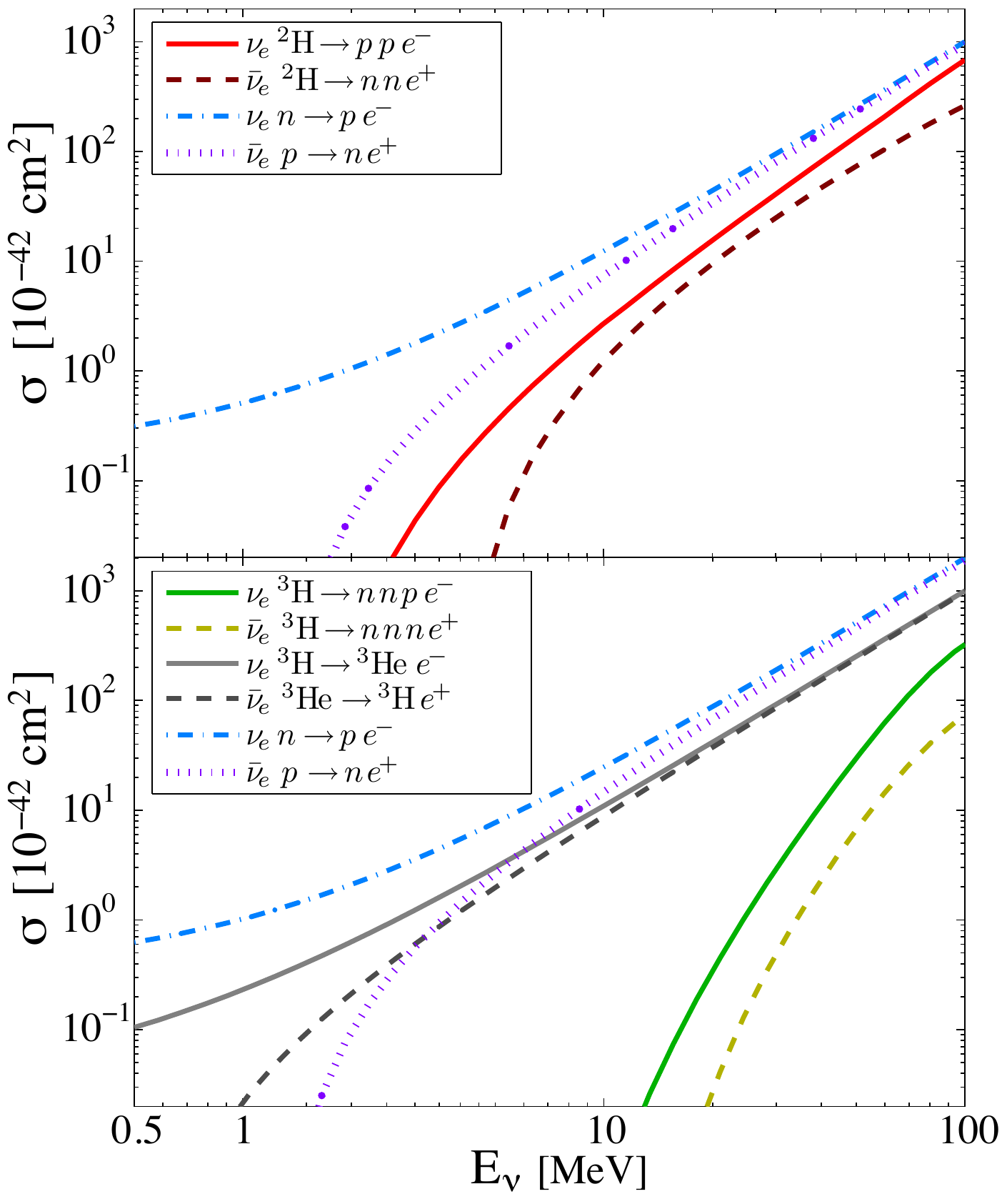}
\caption{Charged current cross sections for $\nu_e$- and $\bar\nu_e$-absorption on light nuclei with $A=2$ (top panel) and $A=3$ (bottom panel), in comparison to those of the Urca processes~\eqref{eq:cc} for charged current reactions.}
\label{fig:light_cc}
\end{figure}
\begin{table}[htp]
\centering
\begin{tabular}{c c}
\hline
\hline
1 & $\nu_e + \, ^2\text{H} \rightleftarrows p + p + e^-$ \\
2 & $\bar\nu_e + \, ^2\text{H} \rightleftarrows n + n + e^+$ \\
3 & $\nu_e + n + n \rightleftarrows \, ^2\text{H} + e^-$ \\
4 & $\bar\nu_e + p  + p \rightleftarrows \, ^2\text{H} + e^+$ \\
5 & $\nu_e + \, ^3\text{H} \rightleftarrows n + p + p + e^-$ \\
6 & $\bar\nu_e + \, ^3\text{H} \rightleftarrows n + n + n + e^+$ \\
7 & $\nu_e + \, ^3\text{H} \rightleftarrows \,^3\text{He} + e^-$ \\
8 & $\bar\nu_e + \, ^3\text{He} \rightleftarrows \,^3\text{H} + e^+$ \\
\hline
9 & $\nu + \, ^2\text{H} \rightleftarrows \, ^2\text{H} + \nu$ \\
10 & $\nu + \, ^3\text{H} \rightleftarrows \, ^3\text{H} + \nu$ \\
11 & $\nu + \, ^3\text{He} \rightleftarrows \, ^3\text{He} + \nu$ \\
12 & $\nu + \, ^2\text{H} \rightleftarrows p + n + \nu$ \\
\hline
\end{tabular}
\caption{Weak processes with light clusters $A=2-3$, separated into spallation (top) and scattering reactions (bottom).}
\label{tab:light}
\end{table}

\subsection{Weak processes with light clusters}
The inclusion of self-consistent weak rates with these light clusters is a much more subtle problem. Unlike for nuclear electron captures~\eqref{eq:e-capture}, where average rates are employed, here rates with individual nuclei must be taken into account. It is common to focus on the most abundant species, $^2$H, $^3$H and $^3$He. In Table~\ref{tab:light} we provide a list of all weak reactions with these light nuclei $A=2-3$, that were considered in the study of \citet{Fischer:2016d}.

Cross sections $\sigma_{\nu^2\text{H}}$ for spallation reactions with $^2$H, (1) and (2) in Table~\ref{tab:light}, are provided by \citet{Nakamura:2001}. They also provide cross sections for inelastic neutrino scattering on deuteron, (12) in Table~\ref{tab:light}. In \citet{Fischer:2016d} it has been realized that these cross sections are related to the electron and positron capture reactions (3) and (4) in Table~\ref{tab:light} via the following replacements,
\begin{eqnarray}
\left(1-f_N\right)\longrightarrow f_N~,\;\;\;\rm and \;\;\;\;\;
\tilde{f}_{^2\text{H}}\longrightarrow\left(1+\tilde{f}_{^2\text{H}}\right)~,
\end{eqnarray}
regarding initial state and final-state phase space occupations of nucleons $N$ and deuteron, as well as the following transformations for the differential cross sections,
\begin{eqnarray}
\frac{d\sigma_{e^-\,^2\text{H}}}{d\Omega_{\nu_e} dp_{\nu_e}}(E)
&\simeq &
\frac{1}{2}
\frac{d\sigma_{\bar\nu_e\,^2\text{H}}}{d\Omega_e dp_e}(E)~,
\\
\frac{d\sigma_{e^+\,^2\text{H}}}{d\Omega_{\bar\nu_e} dp_{\bar\nu_e}}(E)
&\simeq&
\frac{1}{2}
\frac{d\sigma_{\nu_e\,^2\text{H}}}{d\Omega_e dp_e}(E)~,
\end{eqnarray}
assuming relativistic electrons/positrons. Moreover, cross sections for the spallation reactions with $^3$H, (5) and (6) in Table~\ref{tab:light}, were calculated in \citet{Arcones:2008} based on the random phase approximation. In \citet{Fischer:2016d} we realized that the processes (7) and (8) in Table~\ref{tab:light} are significantly more important than the spallation reactions, with three nucleons in the final-state. Cross sections can be given as follows,
\begin{eqnarray}
\sigma_{\nu_e\,^3\text{H}} = \sigma_0 \,\, p_{e^-}E_{e^-} \;,\;\;\;\;\; \sigma_{\bar\nu_e\,^3\text{He}} = \sigma_0 \,\, p_{e^+}E_{e^+} \;,
\label{eq:sigma}
\end{eqnarray}
where
\begin{eqnarray}
\sigma_0 = \frac{G_F^2}{\pi} \frac{V_{ud}^2}{(\hbar c)^4} B(GT) =  1.48\times 10^{-43}~
\frac{\rm cm^2}{\rm MeV^2}~,
\end{eqnarray}
with Fermi constant $G_F$ and $B(GT)=5.97$, known experimentally from the triton decay. Electron and positron energies are related to the $\nu_e$ and $\bar\nu_e$ energies via, $E_{e^-}=E_{\nu_e} + Q_0$ and $E_{e^+}=E_{\bar\nu_e} - Q_0$. The vacuum $Q$-value, $Q_0=0.529$~MeV, is the restmass difference between $^3$He and $^3$H. Fig.~\ref{fig:light_cc} compares the cross sections of all these charged-current weak processes, for $A=2$ (top panel) and $A=3$ (bottom panel) in comparison to those of the Urca processes~\eqref{eq:cc}, where
\begin{eqnarray}
\sigma_0 = \frac{G_F^2}{\pi} \frac{V_{ud}^2}{(\hbar c)^4} (g_V^2+3 g_A^2) =  9.85\times 10^{-44}~\frac{\rm cm^2}{\rm MeV^2}~,
\end{eqnarray}
with vector and axial vector coupling constants $g_V=1.0$ and $g_A=1.26$. Here we relate electron(positron) and $\nu_e$($\bar\nu_e$) energies with the vacuum $Q$-value $Q_0=1.2935$~MeV. 

\begin{figure}[t!]
\includegraphics[width=1\columnwidth]{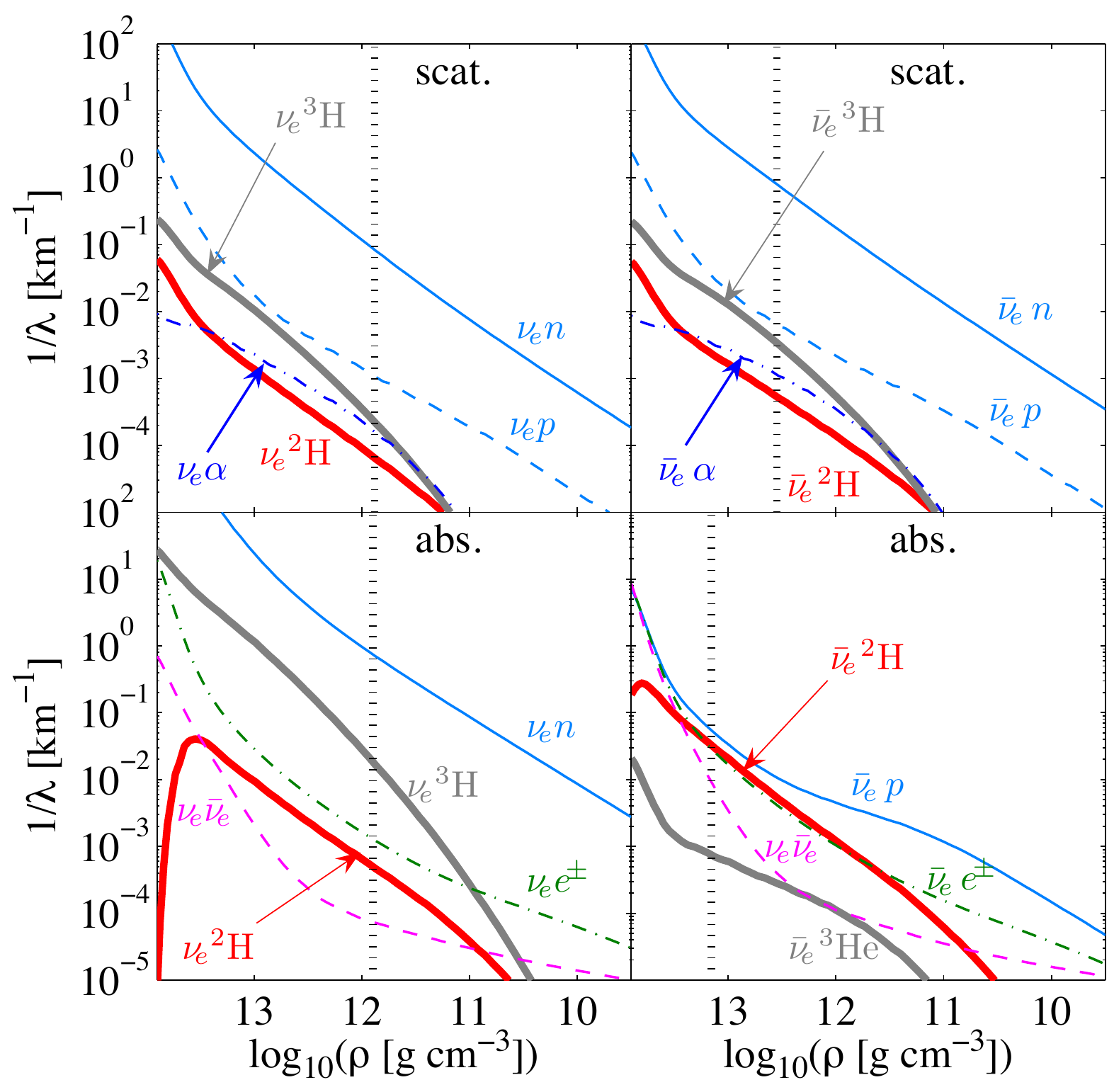}
\caption{Mean-free paths for $\nu_e$ (left panel) and $\bar\nu_e$-reactions (right panel) with light nuclei with $A=2-4$, for neutral-current scattering (top) and charged-current absorption (bottom). The conditions are shown in Fig.~\ref{fig:light_abn}. \citep[Figure adopted from][]{Fischer:2016d}}
\label{fig:mfp}
\end{figure}

When turning these cross sections into reaction rates, the following two aspects are essential: {\em (i)} the vacuum cross sections, including those of \citet{Nakamura:2001}, introduced above have to be ''mapped'' into medium-modified cross sections. The procedure therefore has been introduced in \citet{Fischer:2016d} based on the mean-field treatment with single-particle energies and effective nucleon masses. {\em (ii)} the phase space of the contributing particles has to be taken into account properly \citep[detailed expressions are provided as well in][]{Fischer:2016d}. If this is done accurately, then detailed balance is fulfilled \citep[unlike was done in][]{Furusawa:2013}, and the impact from weak interactions with light nuclear clusters on the supernova neutrino signal and dynamics was found to be negligible in \citet{Fischer:2016d}. The reason for this is illustrated at the example of the mean-free paths in Fig.~\ref{fig:mfp} -- neutral-current scattering (top panel) and the charged-current absorption processes (bottom panel)\footnote{~Mistake in Fig.~(3) of \citet{Fischer:2016d} corrected: labels exchanged $\bar\nu_e\,^3\text{H}<=>\bar\nu_e\,^3\text{He}$ in Fig.~\ref{fig:mfp} bottom panel.}, comparing processes with free nucleons and reactions with light clusters with $A=2-4$. Note that elastic scattering with light clusters is based on the coherent description of \citet{Bruenn:1985en}. The mean free paths, including the neutrino distribution functions, are obtained in detailed core-collapse supernova simulations with Boltzmann neutrino transport of \citet{Fischer:2016d}, including all these reactions with light clusters \citep[for definitions of mean free path as well as neutrinospheres of last scattering, cf.][]{Fischer:2012a}.

For the $\nu_e$, scattering and absorption reactions with free neutrons are the dominating channels by orders of magnitude over those with any light nucleus. This is due to the high abundance of free neutrons in comparison with any other nuclear species (see Fig.~\ref{fig:light_abn}). Moreover, the total opacity is dominated by charged-current absorption on neutrons (see Fig~\ref{fig:mfp}). The situation is similar for $\bar\nu_e$, for which the neutral current channel is also dominated by scattering on neutrons. Nevertheless, $\bar\nu_e$-absorption on protons dominates less strictly over the absorption on $^2$H, however, at increasing density. There, the mean-free path for $\bar\nu_e$-absorption on $^2$H is comparable to other inelastic processes, e.g., $\bar\nu_e$ scattering on electrons and positrons as well as $N$--$N$-bremsstrahlung. Note that in Fig.~\ref{fig:mfp} the labels $\nu_e\bar\nu_e$ correspond to the sum of all pair processes~\eqref{eq:pair}. 

Weak reactions with $A=3$ have in general a negligible role. In Fig.~\ref{fig:mfp} we show only processes~(7) and (8) of Table~\ref{tab:light}, which exceed the break-up reactions with $^3$H by orders of magnitude. The difference to \citet{Arcones:2008} may be due to the lack of final-state blocking contributions. Even though $\nu_e$-absorption on $^3$H exceeds $\nu_e$-absorption on $^2$H, it still lacks short by at least one order of magnitude $\nu_e$-absorption on neutrons. The largely suppressed $\bar\nu_e$-absorption on $^3$He is due to the low abundance of $^3$He. Note also the region of relevance here, illustrated by black vertical dashed lines in Fig.~\ref{fig:mfp}, which mark the locations of the average neutrinospheres of last inelastic (bottom panels) and the effective neutrinospheres (top panels). 

The conditions of Fig.~\ref{fig:mfp} correspond to the early PNS deleptonization phase at about 1~s after the supernova explosion onset has been launched, when the thick layer of accumulated material at the PNS surface from the mass accretion phase falls into the gravitational potential of the PNS. This is the moment of maximum impact of weak processes with light clusters, when the abundance of $^2$H and $^3$H exceed the one of protons maximally. However, the temperatures are already somewhat lower than during the post-bounce mass accretion period prior to the supernova explosion onset. Therefore, \citet{Fischer:2016d} performed supernova simulations based on three-flavor Boltzmann neutrino transport, including in addition to the standard weak processes \eqref{eq:e-capture}--\eqref{eq:ndec} also all weak processes with light clusters with $A=2-3$ shown in Table~\ref{tab:light}. It was found that the impact of weak processes with light clusters on the overall supernova dynamics and the neutrino signal is negligible during the mass accretion phase as well as during the PNS deleptonization phase. 

Finally, when comparing the density domain where nuclear pasta may exists in Fig.~\ref{fig:pasta} with the one where light nuclei are abundant in Fig.~\ref{fig:light_abn}, it becomes evident that both overlap towards high densities. It is therefore important to develop sophisticated models for nuclear pasta that take the presence of light nuclear clusters into account consistently.

\section{Homogeneous matter at supersaturation density}
\label{sec5}
With increasing density, the EOS becomes less and less constrained by nuclear physics. The ab initio approach to the nuclear many-body problem of dilute neutron matter, chiral~EFT, breaks down around normal nuclear matter density $\rho_0$\footnote{~Based on the relativistic mean field model of \citet{Typel:2009sy}, we use $\rho_0=2.6\times 10^{14}$~g~cm$^{-3}$ or equivalent $n_0 = 0.149$~fm$^{-3}$}. It is also well known that (heavy) nuclear clusters dissolve due to the Pauli exclusion principle into homogeneous matter, composed of neutrons and protons \citep[cf.][]{Hempel:2011,Roepke:2013,Furusawa:2017a}. 

\subsection{Excluded volume approach}
In order to explore the uncertainties of the supersaturation density EOS in simulations of core-collapse supernovae, the geometric excluded volume mechanism has been employed in \citet{Fischer:2016a} based on the formalism developed in \citet{Typel:2016}. There, the available volume of the nucleons $V_{N}$ is suppressed via $V_N = V\,\phi$, where $V$ is the total volume of the system. Thereby, the density functional $\phi$ has the following Gaussian type,
\begin{eqnarray}
\label{eq:phi}
\phi(n_{\rm B};\rm v) =
\left\{
\begin{array}{l l}
1 & (n_{\rm B}\leq n_0) \\ & \\
\exp\left\{-\frac{\text{v}\vert \text{v} \vert}{2}\left(n_{\rm B}-n_0\right)^2\right\} & (n_{\rm B}>n_0)
\end{array}
\right.
\end{eqnarray}
for both neutrons and protons, in order to ensure a smooth behavior above saturation density. Based on the choice of the excluded volume parameter, ${\rm v}$, additional stiffening is provided to the EOS for $\rm v>0$ or softening for $\rm v<0$ at supersaturation density. The reference case corresponds to $\rm v=0$, for which the modified NSE approach of \citet{Hempel:2009mc} is selected together with the RMF parametrization DD2 of \citet{Typel:2009sy}, henceforth denoted as HS(DD2), while HS(DD2-EV) denote EOS with excluded volume modifications. Even though DD2 is already rather stiff at supersaturation density, in particular in view of the symmetric matter flow constraint obtained from the detailed analysis of heavy-ion collisions by \citet{Danielewicz:2002}, an even stiffer EOS cannot be ruled out a priori for supernova matter (i.e. at large isospin asymmetry). 

It is important to note that within the approach~\eqref{eq:phi}, nuclear saturation properties remain unmodified, i.e. the saturation density $n_0$ and the symmetry energy at $n_0$ is $J=31.67$~MeV. On the other hand, quantities which relate to derivatives are modified, e.g., the (in)compressibility modulus \citep[a summary of present constraints are given in][]{Stone:2014} varies from $K\simeq 541$~MeV to $K\simeq 201$~MeV for the two selected values of $\rm v=+8.0$~fm$^3$ and $\rm v=-3.0$~fm$^3$ respectively, compared to the reference case $K\simeq 243$~MeV (for $\rm v=0)$. Further details are given in \citet{Typel:2016} and \citet{Fischer:2016a}. 

\begin{figure}[t!]
\includegraphics[width=1\columnwidth]{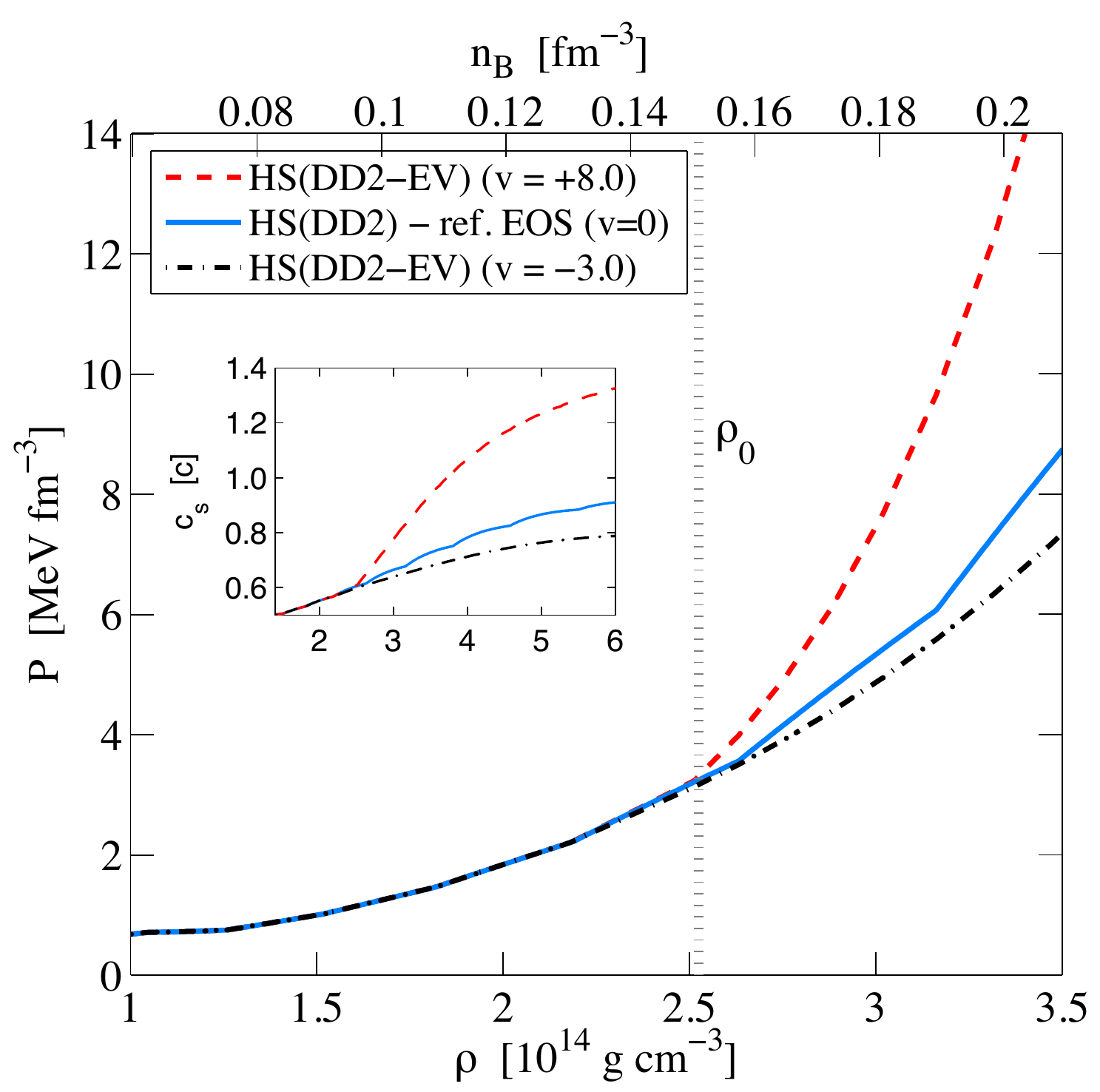}
\caption{High density behavior of the supernova EOS, at selected temperature of $T=5$~MeV and electron fraction $Y_e=0.3$, comparing the reference treatment (v=0) with the modified excluded volume approach with additional stiffening for $\rm v=+8.0$~fm$^{-3}$ and softening for $\rm v=-3.0$~fm$^{-3}$ above supersaturation density ($\rho>\rho_0$). \citep[Figure adopted from][]{Fischer:2016a}.}
\label{fig:eos_ev}
\end{figure}

Fig.~\ref{fig:eos_ev} shows the resulting pressures as a function of density (restmass density $\rho$ at the bottom scale and baryon density $n_{\rm B}$ at the top scale) for the two extreme choices, $\rm v=+8.0$~fm$^{-3}$ and $\rm v=-3.0$~fm$^{-3}$ (units are skipped in the Figure's legend for simplicity). These values are selected such that causality is obtained (see the sound speed in the inlay of Fig.~\ref{fig:eos_ev}) for the relevant supernova densities and maximum neutron star masses are in agreement with the current constraint of about 2~M$_\odot$. Only for $\rm v=+8.0$~fm$^{-3}$ the sound speed exceeds the speed of light, however, at densities far above the central densities reached in the core-collapse simulations which will be further discussed below.

\begin{figure}[t!]
\includegraphics[width=1\columnwidth]{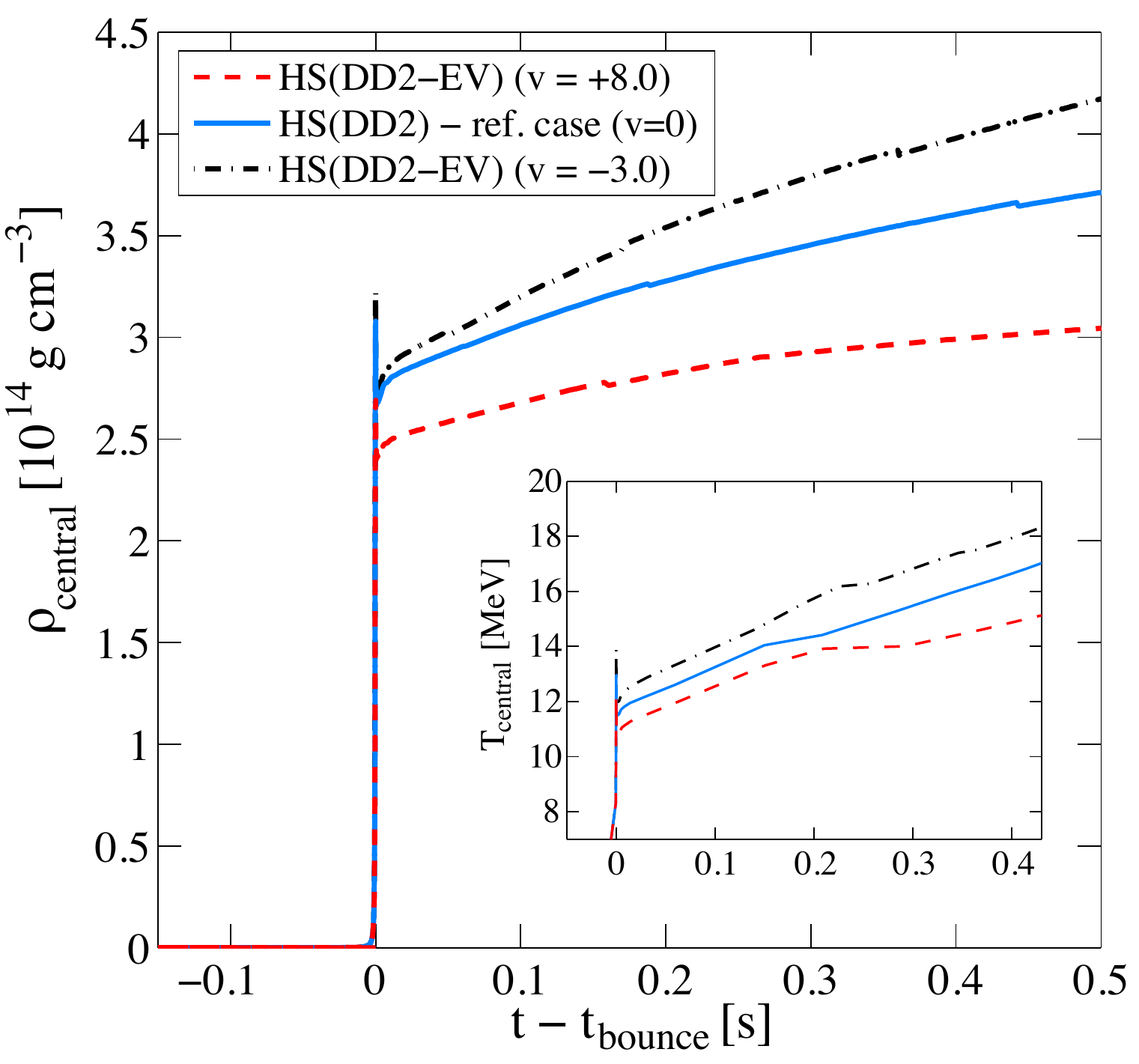}
\caption{Supernova evolution of central density and temperature comparing reference EOS HS(DD2) and variations due the excluded volume approach (see text for details). The dynamical evolution of the reference case is illustrated in Fig.~\ref{fig1} and the neutrino signal is shown in Fig.~\ref{fig:neutrino}. \citep[Figure adopted from][]{Fischer:2016a}}
\label{fig:central_ev}
\end{figure}

The central conditions found during the supernova evolutions are illustrated in Fig.~\ref{fig:central_ev}, in terms of density and temperature. In comparison to the reference case, significantly higher(lower) central densities and temperatures are obtained for the soft(stiff) modification of HS(DD2) with $\rm v=-3.0$~fm$^{-3}$($\rm v=+8.0$~fm$^{-3}$). The central density reaches $4.25\times 10^{14}$~g~cm$^{-3}$ for $\rm v=-3.0$~fm$^{-3}$ in comparison to only $3.0\times 10^{14}$~g~cm$^{-3}$ for $\rm v=+8.0$~fm$^{-3}$, at about 500~ms post bounce. However, it was realized in \citet{Fischer:2016a} that despite such large variation of the central conditions, the overall impact on the supernova dynamics and neutrino signal is negligible \citep[see Figs.~(4) and (5) in][]{Fischer:2016a}. This is because the post-bounce dynamics is determined dominantly at low densities (inhomogeneous nuclear matter, region of high entropy in Fig.~\ref{fig:shellplot}) and the supersaturation density domain of the PNS is generally small, $\sim 0.05-0.1$~M$_\odot$ of the total enclosed baryonic mass of PNS with $\sim1.4-1.8$~M$_\odot$ that result from supernova simulations launched from progenitors in the mass range of $\sim10-30$~M$_\odot$ \citep[cf.][]{Ugliano:2012}.

\subsection{Phase transition to quark matter}
Another similarly uncertain aspect of the EOS at supersaturation density is the question of a possible phase transition from nuclear matter, with hadrons as degrees of freedom, to the deconfined quark gluon plasma with quarks and gluons as the new degrees of freedom. This has long been explored in the context of cold neutron stars. In general, medium properties of quark matter have long been studied \citep[cf.][ and references therein]{Bender:1997jf,Roberts:2000aa,Buballa:2003qv,Alford:2006vz,Klaehn:2007,Blaschke:2017,Pagliara:2007ph,McLerran:2007qj,Sagert:2008ka,Pagliara:2009dg,Blaschke:2009,Klaehn:2010,Klahn:2011fb,Chen:2011my,Fischer:2011,Weissenborn:2012,Bonnano:2012,Blaschke:2013zaa,Klaehn:2013,Kurkela:2014vha,Schaffner:2014}. First-principle calculations of Quantum Chromodynamics (QCD) -- the theory of strongly interacting matter -- are possible by means of conducting large-scale numerical studies \citep[cf.][]{Fodor:2004nz,Ratti:2005jh}, however, only at vanishing density. They predict a smooth cross-over transition from hadronic matter to deconfined quark matter, for a temperature of $T=154\pm9$~MeV at $\mu_B\simeq 0$ \citep[cf.][and references therein]{Katz:2012JHEP,Laermann:2012PRD,Laermann:2012PRL,Katz:2014PhLB}.
Consequently, in astrophysics studies associated with high baryon densities, the two-phase approach is commonly used based on a hadronic EOS and a different EOS for quark matter at high-density. It results in first-order phase transition for which Maxwell or Gibbs constructions are commonly employed. Alternatively, pasta-phases arise when taking into account finite-size effects \citep[][]{Yasutake:2014}. To this end, \citet{Blaschke:2017} studied the appearance of such phases and the associated stability of hybrid stars.

It is important to note that {\em perturbative} QCD is valid only in the limit of asymptotic freedom where quarks are no longer strongly coupled \citep[][]{Kurkela:2014vha}, which automatically excludes astrophysical applications. Hence, in the interior of neutron stars and supernovae, where generally densities are encountered far below the asymptotic limit, effective quark matter models have been commonly employed, e.g., the thermodynamic bag model of \citet{Farhi:1984qu} and models based on the Nambu-Jona-Lasinio (NJL) approach developed by \citet{Nambu:1961tp}, see also \citet{Klevansky:1992qe} and \citet{Buballa:2003qv}.

In simulations of core-collapse supernovae the thermodynamic bag model has been employed in the detailed studies of \citet{Nakazato:2008su} and \citet{Sagert:2008ka}. The latter study assumed low onset densities for the quark-hadron phase transition -- tuned via the bag constant as free parameter -- which are realized in the core of canonical supernovae launched from progenitors with initial masses in the range of 10--20~M$_\odot$. It was found that the thermodynamically unstable region between the stable hadron and quark phases results in the collapse of the PNS, which launches a shock wave that in turn triggers the supernova explosion onset in otherwise non-exploding models in spherical symmetry \citep[for details, see][]{Fischer:2011}. It also releases a millisecond neutrino burst of all neutrino flavors, which was found to be observable with the current generation of neutrino detectors \citep[details can be found in][]{Dasgupta:2009yj}. Unfortunately, these detailed studies violate several 'solid' constraints, e.g., maximum neutron star masses are far below the current value of 2~M$_\odot$, and chiral physics is largely violated. All this urges the need to develop more elaborate phenomenological quark matter EOS, being consistent with current constraints. This marks a major task due to the generally three-dimensional dependencies of supernova matter in terms of temperature, baryon and isospin densities.
Moreover, also the development of weak processes in deconfined quark matter, consistent with the underlying quark-matter EOS, is a major undertaken \citep[for some recent progress, cf.,][and references therein]{Berdermann:2016}.

It has been demonstrated recently in \citet{Klaehn:2015} that the thermodynamic bag model and the NJL approach can be understood as limiting solutions of QCD's in-medium Dyson-Schwinger gap equations \citep[see also][]{Bashir:2012fs,Cloet:2013jya,Chang:2011vu,Roberts:2012sv,Roberts:2000aa,Chen:2008zr,Chen:2011my,Chen:2015mda,Klahn:2009mb}. Based on the newly developed model {\em vBag} of \citet{Klaehn:2015} it has been realized that repulsive vector interaction provides the necessary stiffness for the EOS at high densities in order to yield massive hybrid stars with maximum masses in agreement with the current constraint of $2$~M$_\odot$. Higher-order vector repulsion terms were explored in the non-local NJL model of \citet{Benic:2014jia}. This work has been complemented recently in \citet{Kaltenborn:2017}, providing a relativistic density functional approach to quark matter that implements a confinement mechanism for quarks and allows extensions to finite temperatures. Both aforementioned articles studied the physical case of massive hybrid stars which lead to the 'twin' phenomenon \citep[see also][and references therein]{Haensel:2007yy,Read:2008iy,Zdunik:2012dj,Alford:2013aca}. It relates to the existence of two families of compact stellar objects with similar-to-equal masses but different radii, linked in the mass-radius diagram via a disconnected (i.e. thermodynamically unstable) branch. This interesting situation is part of the science case of the NICER (Neutron Star Interior Composition Explorer)\footnote{~\hyperref[]{https://www.nasa.gov/nicer}} NASA mission, which aims at deducing radii of massive neutron stars at high precision of 0.5~km. See also \citet{Alvarez-Castillo:2016oln} and \citet{Alvarez-Castillo:2017} for a recent Bayesian analysis of constraints on EOS with high-mass twin property using fictitious radius data, that could be provided in the near future, e.g., by NICER and similar missions. Moreover, recent developments point towards a universality condition for the hadron-quark phase transition \citep[cf.][]{Alvarez-Castillo:2016wqj}. 

\begin{figure}[t!]
\includegraphics[width=1\columnwidth]{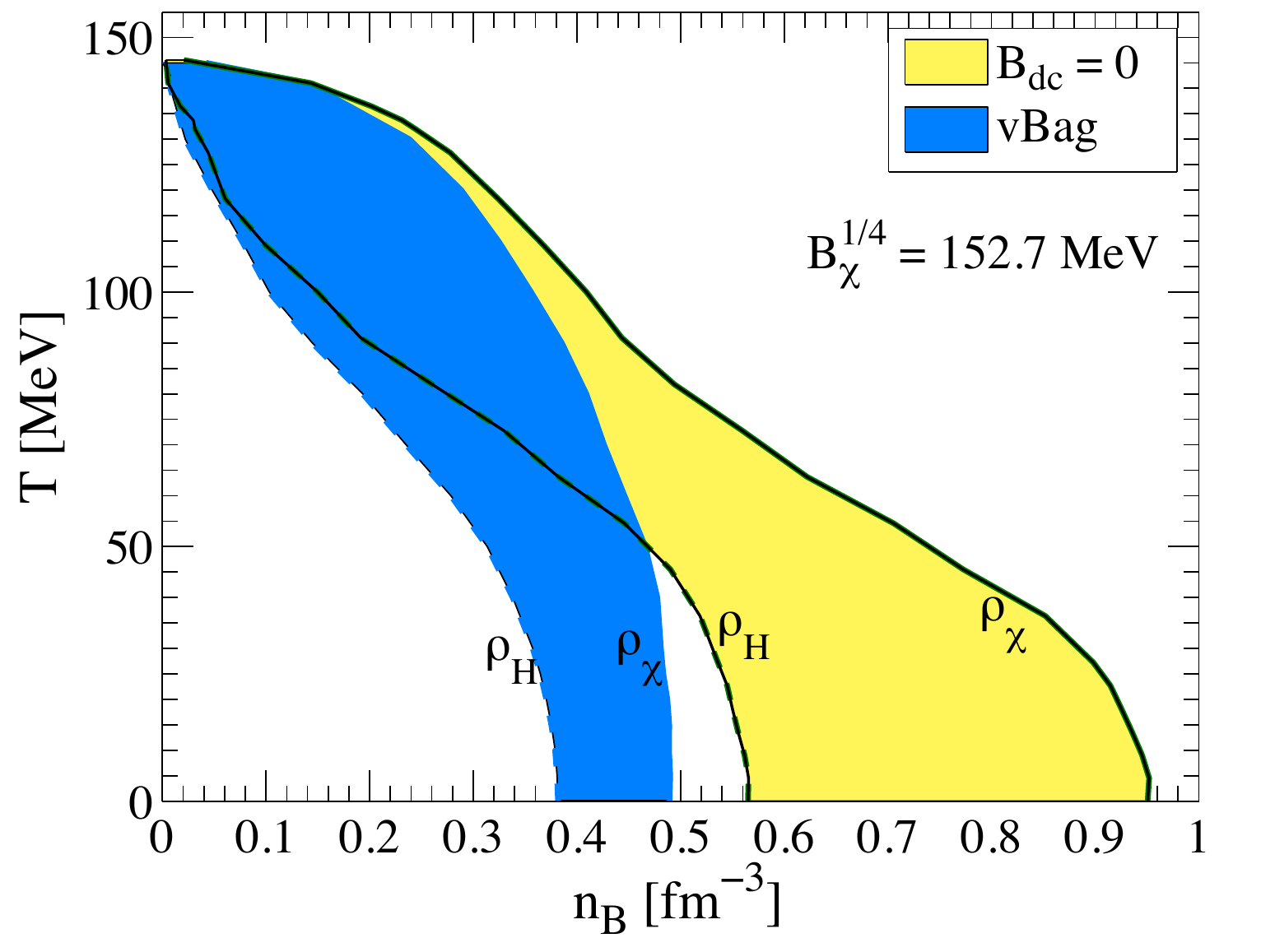}
\caption{vBag phase diagram for matter in $\beta$-equilibrium for the quark matter model developed in \citet{Klaehn:2015}, with the two parameters $B_\chi$ and $B_{\rm dc}$, in comparison to the standard NJL approach ($B_{\rm dc}=0$). \citep[Figure adopted from][]{Klaehn:2017b}}
\label{fig:vbag}
\end{figure}

In \citet{Klaehn:2017a} vBag was extended to finite temperature and arbitrary isospin asymmetry. Fig.~\ref{fig:vbag} illustrates the resulting phase diagram for matter in $\beta$-equilibrium\footnote{~Mistake of \citet{Klaehn:2017b} corrected: labels exchanged $\rm vBag <=> B_{\rm dc}=0$ in the legend of Fig.~\ref{fig:vbag}.}, for a well selected model parameter $B_{\chi}$ (chiral bag constant) which defines chiral symmetry restoration -- marked in Fig.~\ref{fig:vbag} by $\rho_\chi$ -- being in agreement with pion mass and decay constant. Onset densities for the first-order phase transition from the selected underlying hadronic EOS, here HS(DD2), is marked as $\rho_{\rm H}$ in Fig.~\ref{fig:vbag}. Moreover, in contrast to the standard NJL-approach, here (de)confinement is taken into account via a phenomenological parameter ($B_{\rm dc}$) which is determined by the underlying hadronic EOS, being equal to the hadronic pressure at the chiral transition \citep[for details, see][]{Klaehn:2017a}. Hence, it becomes medium dependent, $B_{\rm dc}(T,\mu_Q)$, where $T$ and $\mu_Q$ are temperature and charge chemical potential (the latter referring to isospin asymmetry) respectively. Consequently, additional terms appear which add to the EOS in order to ensure thermodynamic consistency, related to derivatives of $B_{\rm dc}$ with respect to $T$ and $\mu_Q$. 

The canonical approach for constructing phase transitions based on the two-phase approach, i.e. setting $B_{\rm dc}=0$ (also illustrated in Fig.~\ref{fig:vbag}) corresponds to the standard NJL approach commonly used in neutron star studies. It results in significantly higher transition densities for the quark-hadron phase transition. Moreover, this concept leads to the problem that chiral symmetry is (at least partly) restored while matter remains in the confined phase. At present, one can only speculate about the existence of such phase as the currently running heavy-ion collision experiment at NICA \citep[cf. the NICA white paper][and selected contributions therein]{Blaschke:2016} in Dubna (Russia) has not reported any scientific results yet, the low-energy fixed-target experiment at the Relativistic Heavy-Ion Collider (RHIC) in Brookhaven (USA) is expected to produce first results in 2018/2019, and the future Facility for Antiproton and Ion research (FAIR) at the GSI in Darmstadt (Germany) is still under construction. The science runs, in particular experiments related to the compressed baryonic matter physics \citep[cf. the CBM physics book][]{Friman:2011LNP}, will help to shed light onto the phase structure of the entire phase diagram at high densities and lower temperatures.

\section{Summary}
In this review article we summarized selected highlights of EOS developments in simulations of core-collapse supernovae. It covers the entire domain relevant for supernova studies, which can be classified as follows:
\begin{enumerate}
\item {\em Low temperatures}: typically below $T\simeq5$~MeV heavy nuclei exist  -- the treatment of weak processes with heavy nuclei, electron captures, neutrino nucleus scattering and nuclear (de)excitations, are relevant during stellar core collapse. Once heavy nuclei dissociate due to high temperatures as matter is being shock heated early post bounce, weak interactions with heavy nuclei become less important. Towards high densities and low temperatures on the order of few MeV, nuclei become very heavy with $Z\simeq20-35$ and $A\simeq200-400$ (the size of these structures depends on the conditions and the nuclear model). This marks the liquid-gas phase transition at arbitrary isospin asymmetry and in the presence of Coulomb repulsion, where structures shape known collectively as nuclear 'pasta' phases.
\smallskip
\item {\em Intermediate densities and temperatures}: in the domain between $\rho\simeq 10^{9}-10^{13}$~g~cm$^{-3}$ and at $T\simeq 5-10$~MeV, inhomogeneous nuclear matter exists. Here, chiral EFT provides the constraint for dilute neutron matter. However, at arbitrary isospin asymmetry (i.e. with finite abundance of protons) nuclear clusters form. Then, the cluster-virial EOS \citep[cf.][]{Roepke:2013} can provide the constraint at low densities, with particular challenge regarding the extensions to large $A$ \citep[see also][regarding the production of light clusters in heavy-ion collisions]{Bastian:2016}. Inhomogeneous matter corresponds to the shock-heated region at the PNS surface with high entropies per baryon, where in particular light nuclear clusters exist in addition to the unbound nucleons.

\smallskip

{\em Light nuclear clusters} -- play a twofold role, they modify the nuclear EOS in terms of abundances of the free nucleons and single-particle properties {\em and} weak processes with light clusters modify the neutrino transport directly. The latter aspect turns out to be subleading since, taking medium modifications and the particle phase space correctly into account, the leading opacity is due to the Urca and other inelastic processes. This may be slightly modified in multi-dimensional supernova simulations when convection brings high-entropy material down into the cooling layer, which would potentially increase the abundance of light clusters. This will enhance reaction rates with light clusters. However, this remains to be shown in multi-dimensional simulations with neutrino transport which employ the weak processes listed in Table~\ref{tab:light}.
 In any case, it is essential to employ nuclear EOS in supernova simulations that treat the nuclear composition, including all nuclear clusters, consistently. Model EOS with simplified nuclear composition $(n,p,\alpha,\langle A,Z \rangle)$ should be banished from use in supernova studies \citep[for recent efforts, cf.][]{Furusawa:2017b}.
\smallskip
\item {\em At $\rho\gtrsim\rho_0$ and/or high temperatures}: the transition to homogeneous nuclear matter takes place -- composed of free nucleons. The EoS constraints that apply for the state of matter at supersaturation densities are
summarized in \citet{Klaehn:2006}. Of particular importance is the maximum mass of compact stars as predicted by a given EoS model since recent precise measurements of pulsar masses $\sim 2~M_\odot$ allow to exclude EoS which do not possess sufficient stiffness at $\rho\gg\rho_0$. This challenges in particular the appearance of additional degrees of freedom, e.g., strangeness at the hadronic sector in form of hyperons \citep[cf.][]{Sumiyoshi:2009,Bonnano:2012,Weissenborn:2012,Bednarek:2012,Nakazato:2012} as well as the transition to quark matter. Both aspects tend to soften the EOS at $\rho\gg\rho_0$. Moreover, the status of the flow constraint by \citet{Danielewicz:2002} is debated. It is desireable to perform direct comparisons with data on directed and elliptic flow from heavy-ion collision experiments for given model EOS. To this end, advanced simulation programs have to be employed. These would apply in particular in the low-energy region starting from SIS and AGS energies to the NICA and FAIR domain. The recently developed event simulation program THESEUS which is based on a three-fluid hydrodynamics model, provides an appropriate tool for studying these questions and for comparing flow patterns of EOS with and without QCD phase transitions \citep[see][]{Batyuk:2016qmb}. 
%
\end{enumerate}
A detailed review of the presently available EOS has been provided in \citet{Oertel:2017}, with a slightly different focus. However, some of the aspects discussed here are being explored in \citet{Oertel:2017} as well.

\begin{acknowledgements}
The authors acknowledge support from the Polish National Science Center (NCN) under grant numbers UMO-2016/23/B/ST2/00720~(TF), UMO-2011/02/A/ST2/00306~(DB) and UMO-2014/13/B/ST9/02621~(NUB and MC). DB and GR acknowledge support by the MEPhI Academic Excellence Project under contract No.~02.a03.21.0005. WN was supported by the Research Corporation for Science Advancement through the Cottrell College Science Award no. 22741. GMP and ST are partially supported by the DFG through grant SFB1245. This work was supported in part by the COST Actions MP1304 ``NewCompStar", CA15213 ``THOR'' and CA16117 ``ChETEC''.
\end{acknowledgements}

\begin{appendix}
\section{Appendix: Neutrino heating and cooling rates}
\label{appendix}
For $\nu_e$ from electron capture, the change in the entropy per nucleon, $s$, of matter with temperature $T$ is given in \citet{Bruenn:1985en} as follows\footnote{~Here we are using the same notation as in Fig~\ref{fig:heating}, i.e. 'cc', 'pair' and '$\nu e^\pm$'.},
\begin{equation}
\label{eq:dsec}
T \frac{ds}{dt} = \left.\frac{dQ_{\rm cc}}{dt}\right\vert_{\rm net} - (\mu_e + \mu_p - \mu_n) \frac{dY_e}{dt}~, 
\end{equation}
with chemical potentials of electrons $\mu_e$, protons $\mu_p$ and neutrons $\mu_n$. In Eq.~\eqref{eq:dsec} the last term accounts for the change in composition due to electron captures and has been determined assuming thermodynamical equilibrium. $dQ_{\rm cc}/dt\vert_{\rm net}$ is the net energy exchange rate between neutrinos and matter, defined to be negative  when neutrino emission dominates. It can be determined from the electron neutrino emissivity, $j(E)$, and opacity, $\chi(E)$, for nuclear electron captures~\eqref{eq:e-capture} as well as the Urca processes~\eqref{eq:cc} according to \citet{Bruenn:1985en} and \citet{Mezzacappa:1993gm} as follows,
\begin{align}
\label{eq:dqdt_ve}
&
\left.\frac{dQ_{\rm cc}}{dt}\right\vert_{\rm net} = \frac{2\pi}{n_{\rm B}}
\int dE E^3 d(\cos\theta)
\\
&
\;\;\;\;\times\;
\biggr\{ \chi(E) f_{\nu_e} (E,\theta) - j(E) \left(1-f_{\nu_e}(E,\theta)\right)\biggr\}~,
\nonumber
\end{align}
with baryon density $n_{\rm B}$. Expression~\eqref{eq:dqdt_ve} takes into account the full phase-space dependence of all contributing particles, including final-state blocking of $\nu_e$ expressed here via the phase-space distribution function $f_{\nu_e}$ depending on neutrino energy $E$ and lateral momentum angle $\theta$. For a definition of the neutrino phase-space setup, see Fig.~(1) in \citet{Mezzacappa:1993gm}.

For neutrino-pair production~\eqref{eq:pair} we have,
\begin{equation}
\label{eq:dspair}
T \frac{ds}{dt} = \left.\frac{dQ_{\rm pair}}{dt}\right\vert_{\rm net}~,
\end{equation}
where $dQ_{\rm pair}/dt\vert_{\rm net}$ is determined from the neutrino-pair emission and absorption kernels, $\mathcal{R}^{\rm emi/abs}$, as follows,
%
\begin{align}
\label{eq:qnetpair}
&
\left.\frac{dQ_{\rm pair}}{dt}\right\vert_{\rm net}
=
\frac{(2\pi)^2}{n_{\rm B}}
\int dE E^2 d(\cos\theta) \, d\bar E \bar E^2 d(\cos\bar\theta)
\\
&\;\;\;\;\;\times
(E+\bar E)\biggr\{
f_{\nu}(E,\theta) f_{\bar\nu}(\bar E,\bar\theta)\,\mathcal{R}^{\rm abs}(E+\bar E,\theta,\bar\theta)
\nonumber
\\
&\;\;\;\;\;-
\left(1-f_{\nu}(E,\theta)\right)\left(1-f_{\bar\nu}(\bar E,\bar\theta)\right)\mathcal{R}^{\text{emi}}(E+\bar E,\theta,\bar\theta)\biggr\}~,
\nonumber
\end{align}
%
taking into account the corresponding neutron and antineutrino phase-space occupations, $f_\nu(E,\theta)$ and $f_{\bar\nu}(\bar E, \bar\theta)$ respectively.

Finally, for inelastic $\nu$ scattering with electrons and positrons~\eqref{eq:scat2} we have,
\begin{equation}
\label{eq:dsnue}
T \frac{ds}{dt} = \left.\frac{dQ_{\nu e^\pm}}{dt}\right\vert_{\rm net}~,
\end{equation}
with $dQ_{\nu e^\pm}/dt\vert_{\rm net}$ being related to the $\nu$ scattering kernels by a similar expression to equation~\eqref{eq:qnetpair}.

\end{appendix}

\bibliographystyle{pasa-mnras}
\bibliography{references}

\end{document}